\pdfoutput=1
\documentclass[11pt,twoside,a4paper]{article}

\usepackage[utf8]{inputenc}
\usepackage[sort&compress,numbers,colon,merge]{natbib}
\newcommand{\cAfour}{\cite{Babu:2005uq,*Ma:2004qy,*Babu:2003fk,*Ma:2001lr,*He:2006fj,*Altarelli:2006qy,Altarelli:2005uq}}

\newcommand{\prd}{Physical Review D.}

\newcommand{\cSfour}{\cite{Pakvasa:1978tx,*Yamanaka:1981pa,*Brown:1984mq,*Brown:1984dk,* Lee:1994qx,*Ma:2005pd,*Hagedorn:2006ug,*Cai:2006mf,*Caravaglios:2006aq,*Zhang:2006fv,*Koide:2007sr,*Parida:2008pu,*Bazzocchi:2008ej,*Ishimori:2008fi,*Bazzocchi:2009da,*Altarelli:2009gn,*Ishimori:2009ns,*Grimus:2009pg,*Ding:2009iy,*Meloni:2009cz,*Morisi:2010rk,*Dutta:2009bj,Lam:2008sh}}
\newcommand{\cTseven}{\cite{Luhn:2007fj}}

\newcommand{\cTprime}{\cite{Mohapatra:2004mz,*Ding:2008zr,*Frampton:2009fr,*Frampton:2007ys,*Aranda:2007rt,*Carr:2007vn,*Feruglio:2007yq,*Chen:2007kx}}

\newcommand{\cDeltatwentyseven}{\cite{Bazzocchi:2009qf,*Luhn:2007ul,*Grimus:2008ve,*de-Medeiros-Varzielas:2007gf}}
\newcommand{\cAfourVEV}{\cite{Altarelli:2005yx,He:2006dk}}
\newcommand{\cDifferentSubgroups}{\cite{Altarelli:2005yx,He:2006dk}}
\newcommand{\cHPS}{\cite{Harrison:1999cf,*Harrison:2002er,*Harrison:2002kp}}

\usepackage{a4wide}
\usepackage{fancyhdr}
\usepackage{graphicx}
\usepackage{url}
\usepackage{amsfonts}
\usepackage{amsmath,amssymb,amsthm}

\usepackage{fmdg}
\usepackage{mathrsfs}
\usepackage{subfigure}
\usepackage{multirow}
\usepackage{color}
\usepackage[colorlinks]{hyperref}
\usepackage{cancel}
\usepackage{bbm}
\hypersetup{
pdfauthor={Martin Holthausen and Michael A. Schmidt},
pdftitle={Natural Vacuum Alignment from Group Theory: The Minimal Case}
}

\makeatletter
\renewcommand{\fnum@table}{\textbf{\tablename~\thetable}}
\renewcommand{\fnum@figure}{\textbf{\figurename~\thefigure}}
\makeatother

\newcommand{\diag}{\ensuremath{\mathrm{diag}}}
\newcommand{\I}{\ensuremath{\mathrm{i}}}

\newcommand{\eV}{\ensuremath{\,\mathrm{eV}}}

\newcommand{\GeV}{\ensuremath{\,\mathrm{GeV}}}

\newcommand{\braket}[1]{\ensuremath{\left<#1\right>}}
\newcommand{\ev}[1]{\ensuremath{\left\langle#1\right\rangle}}
\newcommand{\hc}{\ensuremath{\text{h.c.}}}
\newcommand{\Order}[1]{\ensuremath{\mathcal{O}(#1)}}
\newcommand{\be}{\begin{equation}}
\newcommand{\ee}{\end{equation}}
\newcommand{\ba}{\begin{eqnarray}}
\newcommand{\ea}{\end{eqnarray}}
\newcommand{\SU}[1]{\ensuremath{\mathrm{SU}(#1)}}
\newcommand{\U}[1]{\ensuremath{\mathrm{U}(#1)}}
\newcommand{\SO}[1]{\ensuremath{\mathrm{SO}(#1)}}

\newcommand{\Eqref}[1]{Eq.~(\ref{#1})}

\newcommand{\Figref}[1]{Fig.~\ref{#1}}
\newcommand{\Tabref}[1]{Tab.~\ref{#1}}
\newcommand{\Secref}[1]{sec.~\ref{#1}}

\renewcommand{\subsubsection}[1]{\vspace{1ex}\mathversion{bold}{\bf #1:}\mathversion{normal}}

\newcommand{\Rep}[1]{\ensuremath{\underline{\mbox{\textbf{#1}}}}}
\newcommand{\MoreRep}[2]{\ensuremath{\underline{\mbox{\textbf{#1}}} _{\mbox{\textbf{#2}}}}}

\newcommand{\GAP}{\texttt{GAP}}
\newcommand{\Repsn}{\texttt{REPSN}}

\newcommand{\Discrete}{\texttt{Discrete}}
\newcommand{\SmallGroups}{\texttt{SmallGroups}}

\newcommand{\abs}[1]{\ensuremath{\left\vert#1\right\vert}}

\begin{document}
\allowdisplaybreaks[1]


\begin{titlepage}

\ \vspace*{-15mm}
\begin{flushright}
IPPP/11/62\\
DCPT/11/124
\end{flushright}
\vspace*{5mm}

\begin{center}
{\Huge\sffamily\bfseries 
Natural Vacuum Alignment from Group Theory: The Minimal Case
}
\\[10mm]
{\large
Martin Holthausen\footnote{\texttt{martin.holthausen@mpi-hd.mpg.de}}$^{(a)}$ and
Michael A.~Schmidt\footnote{\texttt{michael.schmidt@unimelb.edu.au}}$^{(b)(c)}$}
\\[5mm]
{\small\textit{$^{(a)}$
Max-Planck Institut f\"ur Kernphysik, Saupfercheckweg 1, 69117
Heidelberg, Germany
}}\\
{\small\textit{$^{(b)}$
Institute for Particle Physics Phenomenology (IPPP), University of Durham,
Durham DH1 3LE, UK
}}\\
{\small\textit{$^{(c)}$
ARC Centre of Excellence for Particle Physics at the Terascale,
School of Physics, The University of Melbourne, Victoria 3010, Australia
}}

\end{center}
\vspace*{1.0cm}
\date{\today}

\begin{abstract}
\noindent
Discrete flavour symmetries have been proven successful in explaining
the leptonic flavour structure. To account for the observed mixing pattern, the flavour symmetry has to be broken to different subgroups in the
charged and neutral lepton sector.
 However, cross-couplings via non-trivial contractions in the scalar potential force the group to break to the same subgroup. We present a solution to this problem by extending the flavour group in such a way that it preserves the flavour structure, but leads to an 'accidental' symmetry in the flavon potential.

We have searched for symmetry groups up to order 1000, which forbid all dangerous cross-couplings and extend one of the interesting groups $A_4$, $T_7$, $S_4$, $T^\prime$ or $\Delta(27)$.
We have found a number of candidate groups and present a model based on one of the smallest extensions of $A_4$, namely $Q_8\rtimes A_4$. We show that the most general nonsupersymmetric potential allows for the correct vacuum alignment.  We investigate the effects of higher dimensional operators on the vacuum configuration and mixing angles, and give a see-saw-like UV completion. Finally, we discuss the supersymmetrization of the model.
Additionally, we release the Mathematica package \Discrete\ providing various useful tools for model building such as easily calculating invariants of discrete groups and flavon potentials.
\end{abstract}

\end{titlepage}

\newpage
\setcounter{footnote}{0}

\section{Introduction}
Over the last decade, neutrino oscillation experiments have measured the mixing angles of the leptonic mixing matrix $U_{PMNS}$ to quite some accuracy~\cite{Schwetz:2011kx,*Schwetz:2011uq, Fogli:2011qn, GonzalezGarcia:2010er}. It turns out that two of the mixing angles, namely the solar and  atmospheric angles  $\theta_{12}$ and $\theta_{23}$, are large while the third one, the reactor angle $\theta_{13}$, is small. Recently, there has been a hint of a non-vanishing third mixing angle $\theta_{13}$ by the T2K experiment~\cite{Abe:2011sj} close to the upper bound of the CHOOZ experiment~\cite{Apollonio:2002gd}. Remarkably, the current best fit values (taking into account the recent measurements) for the case of normal neutrino mass hierarchy~\cite{Schwetz:2011kx,*Schwetz:2011uq} 
\begin{align*}
\sin^2 \theta_{12}&=0.312^{+ 0.017}_{- 0.015},&
\sin^2 \theta_{23}&=0.52^{+0.06}_{-0.07}&\mathrm{and}&&
\sin^2 \theta_{13}&=0.013^{+ 0.007}_{- 0.006}
\end{align*}
are rather close to tri-bimaximal mixing
\begin{align*}
\sin^2\theta_{12}&=\frac13 , &
\sin^2\theta_{23}&=\frac12&\mathrm{and}&& 
\sin^2\theta_{13}&=0\;,
\end{align*}
which was first proposed by Harrison, Perkins and Scott~\cHPS. 
Before the hint of a non-vanishing $\theta_{13}$ mixing angle by T2K and MINOS~\cite{MINOS-Collaboration:2011fk}, the description by tri-bimaximal mixing was even better.

This interesting observation has led many authors
to search for a theoretical explanation of this curious mixing
pattern. While in the quark sector the mixing angles are small and can
be explained by a Froggatt-Nielsen type U(1) symmetry~\cite{Froggatt:1978nt}, which also
accounts for the quark mass hierarchies, in the neutrino sector so far
the most fruitful approach has been to introduce a non-abelian
discrete symmetry, like $A_4$~\cAfour, $T_7$~\cTseven, $S_4$~\cSfour, $T^\prime$~\cTprime~and $\Delta(27)$~\cDeltatwentyseven~among others. 
The left-handed lepton doublets are commonly assigned to a non-trivial representation of the flavour group, e.g.~a triplet.
Subsequently, this symmetry is spontaneously broken to different subgroups in the charged lepton and neutrino sector~\cDifferentSubgroups. This requires at least two scalar fields to obtain vacuum expectation values (VEVs), which are pointing in two different directions in the space of the flavour symmetry. This is commonly denoted by VEV alignment, which we are going to address in this article.

The most well-studied non-abelian discrete group is the group $A_4$, which is the symmetry group of a regular tetrahedron. It is the smallest group with an irreducible three dimensional representation $\Rep{3}$.
If one assigns the lepton doublets as $L\sim\Rep{3}$ and charged
leptons to the three singlet representations $e^c$, $\mu^c$, $\tau^c$
as $\MoreRep{1}{1}$, $\MoreRep{1}{2}$ and $\MoreRep{1}{3}$, the
tri-bimaximal mixing structure is generated due to a mismatch of the
vacuum expectation values of the flavons $\chi\sim\Rep{3}$ and
$\phi\sim\Rep{3}$ that couple to charged leptons and the neutrinos,
respectively. If $A_4$ is broken down to the subgroup $Z_3$ in the
charged lepton sector, e.g. by $\ev{\chi}\propto (1,1,1)^T$, and it is broken
to the subgroup $Z_2$ in the neutrino sector, e.g. by $\ev{\phi}\propto (1,0,0)^T$, 
the unitary transformation that connects the most general mass matrices invariant under these symmetries is given by
\begin{align}
U_{PMNS}&=U_{HPS} U_{23}(\theta)\quad\quad\mathrm{with}\quad\quad
U_{HPS}=\left(
\begin{array}{ccc}\sqrt{\frac{2}{3}}&\frac{1}{\sqrt{3}}&0\\-\frac{1}{\sqrt{6}}&\frac{1}{\sqrt{3}}&-\frac{1}{\sqrt{2}}\\\frac{1}{\sqrt{6}}&\frac{1}{\sqrt{3}}&\frac{1}{\sqrt{2}} \end{array}\right)
\label{eq:HPSmatrix}
\end{align}
where $U_{HPS}$ denotes the tri-bimaximal mixing matrix and $U_{23}(\theta)$
denotes a $2-3$ rotation by the angle $\theta$, which is generated by operators of the form $LL(y\,\eta-\hc)$ with $\eta\sim\MoreRep{1}{2,3}$. In case, these operators do not contribute to the neutrino mass matrix, i.e. $\mathrm{Im}\left(y\,\ev{\eta}\right)=0$, the leptonic mixing matrix is given by the tri-bimaximal mixing matrix $U_{HPS}$. (In light of the recent hint on a non-vanishing $\theta_{13}$, a non-vanishing operator of this type has been discussed in~\cite{Shimizu:2011xg}.) 

The prediction of tri-bimaximal mixing in $A_4$ models thus
requires a special vacuum alignment\footnote{See~\cite{Toorop:2011jn} for a model that can accommodate the large value for $\theta_{13}$ suggested by T2K and still needs a special vacuum alignment.}, which should have a dynamical
origin within the model. In the most straightforward dynamical model, namely the usual scalar potential, the cross coupling terms connecting $\phi$ and $\chi$ via non-trivial $A_4$ contractions, e.g. $(\chi^2)_{\MoreRep{1}{2}} (\phi^2)_{\MoreRep{1}{3}}$,  forbid the desired vacuum alignment\cAfourVEV, as will be reviewed in the next section. This vacuum alignment problem is not limited to $A_4$ models, but a general problem of most of the symmetry
groups, that have been studied.
In the literature, several mechanisms have been proposed to address this problem:
\begin{enumerate}
\item[(i)]
in models with extra dimensions, it is possible to localize
 the two flavons 
differently in the extra dimensions with no (or negligible) overlap of the wave functions, thereby forbidding (or suppressing) all cross-couplings~(see e.g.~\cite{Altarelli:2005uq} for a localization on different branes)~\footnote{Models with extra dimensions also allow an explicit breaking of the flavour symmetry via boundary conditions~\cite{Kobayashi:2008ih}.},
\item[(ii)]
in a supersymmetric framework it is possible to introduce $R$ charges and driving fields with $R$ charge $2$ such that the tadpole of the driving fields forces the desired vacuum alignment (see e.g.~\cite{Altarelli:2005yx}).
\end{enumerate}
While these approaches are interesting and worth studying, we will develop further an idea by Babu and Gabriel~\cite{Babu:2010bx}, who have
suggested a group-theoretical mechanism to forbid the dangerous
cross-couplings and thus a way to realise the VEV alignment without using $R$-symmetries in supersymmetry or brane constructions.

They proposed an extension of the flavour group $A_4$ in such a way that the Standard Model leptons only transform under the $A_4$ subgroup of the full flavour group. In the scalar sector, the flavon $\chi$ of the charged lepton sector also transforms only under the $A_4$ subgroup, while the flavon $\phi$ of the neutrino sector transforms under the full flavour group $G$. For a suitably chosen group $G$, it is then possible that the additional group transformations forbid the contractions $(\phi\phi)_{\MoreRep{1}{2,3}} $ and $(\phi\phi)_{\MoreRep{3}{1}} $, which lead to the dangerous couplings in the scalar potential and make the correct vacuum alignment impossible. In other words, the additional discrete symmetry leads to an accidental symmetry at the renormalizable level in the flavon potential, $G\times A_4$, allowing for a different breaking of the two $A_4$ subgroups of the accidental symmetry. The coupling to leptons only respects the diagonal $A_4$ subgroup, which is thus broken to different subgroups in the charged lepton and neutrino sectors, as desired. 

Note that this construction requires that the additional group
generators cannot all commute with the generators of $A_4$, i.e. the
flavour group $G$ cannot be a direct product of $A_4$ with some other
group, but it has to be a slightly more general object, a
\emph{semidirect product}. In fact, we will generalize this
construction one step further and look into more general group
extensions. This will be explained in more detail in \Secref{sec:GroupScan}.

In their work, Babu and Gabriel used a special type of semidirect product, a so-called \emph{wreath product}
 of $A_4$ with $S_3$, i.e. the product of four factors of $S_3$ which are evenly permuted by the group $A_4$. It is thus a very complicated flavour group of order $12\cdot 6^4=15552$ and requires the use of very large representations up to dimension $48$. This model further suffers from a fine-tuning problem, as the diagonal and off-diagonal elements of the neutrino mass matrix are generated by operators with very different mass dimension, 5 and 10, even though both entries should be of comparable size.

In this article, we are addressing these issues and are presenting the
result of a search for simpler and more attractive semidirect product
groups $G=N\rtimes H$ as well as general group extensions $G$ satisfying
$G/N\cong H$ with $H$ being $A_4$, $T_7$, $S_4$, $T^\prime$ or $\Delta(27)$~\footnote{This mechanism is of course not limited to these five groups, but is also relevant for other flavour groups, e.g. larger groups constructed to be in agreement with the recent T2K measurement such as~\cite{Toorop:2011jn}.}, which lead to an accidental symmetry $G\times H$ in the flavon potential at the renormalizable level. We included all discrete groups up to order $1000$ in our search and found several candidate groups. The smallest candidate groups are of order $96$, in particular the semidirect product group of the quaternion group with $A_4 $, $Q_8\rtimes A_4$, which we discuss in more detail in \Secref{sec:SmallestGroup}. This group does not have representations of size larger than four and, in the model we present, on-and-off-diagonal entries in the neutrino mass matrix are generated at the same order.

This work has been accompanied by the development of the Mathematica package \Discrete\ facilitating the calculation of the different covariants of a discrete flavour group. Dirac and Majorana mass matrices as well as the flavon potential can be calculated automatically up to an arbitrary order. 
It can access the large group catalogues implemented in \GAP~\cite{GAP4:2011}. Its \SmallGroups~\cite{SmallGroups:2011} catalogue, for example, contains all discrete groups up to order $2000$ with the exception of order $1024$. 

 The outline of the paper is as follows. In \Secref{sec:VEValignment},
 we discuss the VEV alignment problem in the context of $A_4$. Our
 search for semidirect product groups and the results are described
 in \Secref{sec:GroupScan}.  Readers, who are not interested in the technical details of
 the construction, may skip this section. In \Secref{sec:SmallestGroup}, we discuss
 the smallest candidate group $Q_8\rtimes A_4$ and construct a
 model using it. It is the minimal model based on $A_4$ allowing for the correct vacuum alignment. Higher order corrections are discussed in
 \Secref{sec:higherorderop}. An ultraviolet (UV) completion is
 presented in \Secref{sec:UV-completion} and a supersymmetric
 version is given in \Secref{sec:SUSY-Model}. The Mathematica package
 \Discrete\ is introduced in \Secref{sec:Mathematica}.
Finally, we conclude in \Secref{sec:Conclusions}. Group-theoretical
details are summarised in the appendix.


\section{\mathversion{bold}VEV Alignment in $A_4$ Revisited}
\label{sec:VEValignment}

In this section, we want to remind the reader about the difficulties
one encounters when minimising flavon potentials~\cAfourVEV. Here we
focus on the problem one faces in the most straightforward case -- namely the case of a nonsupersymmetric scalar potential. The case of softly-broken supersymmetry is included in this analysis as SUSY only further restricts the dimensionless couplings of the potential while care has to be taken not to have flat directions in the cubic superpotential. We will come back to the SUSY case in \Secref{sec:SUSY-Model}.

For simplicity, we consider $A_4$, the symmetry group of the tetrahedron, which is the smallest discrete group with a three dimensional irreducible representation. It is presented by $\braket{S,T\vert S^2=T^3=(ST)^3=1}$. As we have discussed in the introduction, tri-bimaximal mixing is generated by breaking this group to its subgroups generated by $S$ and $T$ in the neutrino and charged lepton sectors, respectively. The character table and the representation matrices for the three dimensional representation are given in Table \ref{tab:A4representations}.

\begin{table}
\centering
\subtable[Character Table]{
\begin{tabular}{|l|cccc|}\hline
& 1 & $T$ & $T^2$ & $S$\\\hline 
\MoreRep{1}{1} & 1 &1 &1 &1\\
\MoreRep{1}{2} & 1 & $\omega$ & $\omega^2$ & 1\\
\MoreRep{1}{3} & 1 & $\omega^2$ & $\omega$ & 1\\
\Rep{3} & 3 & 0 & 0 & -1\\\hline
\end{tabular}
}
\hspace{1cm}
\subtable[Generators]{\begin{minipage}{6cm}
\begin{equation*}
S=\left(\begin{array}{ccc}
1&0&0\\
0&-1&0\\
0&0&-1
\end{array}\right) \quad\quad
T=\left(\begin{array}{ccc}
0&1&0\\
0&0&1\\
1&0&0
\end{array}\right)
\end{equation*}
\end{minipage}}
\hspace{1cm}
\caption{Character table of $A_4$ as well as matrix representation of generators in \Rep{3}. Here $\omega=e^{\I 2 \pi/3}$.\label{tab:A4representations}}
\end{table}

Let us look at the potential\footnote{Here, we have assumed a discrete symmetry ($\chi\rightarrow -\chi, \, \, f^c \rightarrow - f^c$, with $f=e,\mu,\tau$) that separates the charged lepton from the neutrino sector, as it is common practice. The operator $(\chi \chi)_{\MoreRep{3}{1}} \cdot (\chi \chi)_{\MoreRep{3}{1}}$, which one would naively expect, can be expressed as a linear combination of the other operators.}
\begin{equation}
V_\chi=m_{0}^2\, (\chi \chi)_{\MoreRep{1}{1}}
+\lambda_1\, (\chi \chi)_{\MoreRep{1}{1}}(\chi \chi)_{\MoreRep{1}{1}}+\lambda_2\, (\chi \chi)_{\MoreRep{1}{2}}(\chi \chi)_{\MoreRep{1}{3}}
\end{equation}
 of a real scalar triplet  $\chi$ of $A_4$ that couples to charged leptons via the operators~ $\ell \tilde{H} f^c \chi /\Lambda$  
and should therefore acquire a VEV $\ev{\chi}=\left(v^\prime,\,v^\prime,\,v^\prime\right)^T$ conserving the $Z_3$ subgroup generated by T. The symmetry breaking of $A_4$ to $\langle S \rangle\cong Z_2$ (which
might or might not be due to the VEV of another triplet $\phi$ in the
neutrino sector with $\ev{\phi}=\left(w,\,0,\,0\right)^T$), will lead to
the following soft terms in the potential:
\begin{align}
V_{soft,Z_2}=m_A^2 \chi_1^2+m_B^2 \chi_2^2+m_C^2 \chi_2 \chi_3
\end{align}
The minimisation conditions of the full potential $V=V_\chi+V_{soft,Z_2}$ 
evaluated at the desired minimum $\ev{\chi}=\left(v^\prime,\,v^\prime,\,v^\prime\right)$ result in 
\begin{subequations}
\begin{align}
0&=\left[\frac{\partial V}{\partial\chi_1}\right]_{\chi_i=v^\prime}=\frac{2}{\sqrt{3}} \left(m_0^2+ \sqrt{3} m_A^2\right)\,v^\prime+4 \lambda _1 {v^\prime}^3\\
0&=\left[\frac{\partial}{\partial \chi_2}V-\frac{\partial}{\partial \chi_3}V\right]_{\chi_i=v^\prime}=2 \, m_{B}^2\,v^\prime\\
0&=\left[\frac{\partial}{\partial \chi_1}V-\frac{\partial}{\partial \chi_3}V\right]_{\chi_i=v^\prime}=\left( 2 \, m_{A}^2 - m_C^2\right) \,v^\prime
\end{align}
\end{subequations}
The vacuum alignment thus requires $m_{B}^2=0$ and $m_C^2=2 m_{A}^2$ and therefore two completely different $A_4$ contractions need to have the same coupling in the scalar potential, an option we exclude as fine-tuning. Even if one sets the terms $m^2_{A,B,C}$ to zero, they will still be generated on loop-level and disturb the VEV alignment. The breaking of $A_4$ to two different subgroups thus requires a systematic mechanism to forbid $m^2_{A,B,C}$. 

On the contrary, soft breaking terms which preserve the same $Z_3$ subgroup 
\begin{equation}
V_{soft,Z_3}= m_s^2\, \left(\chi_2\chi_3+\chi_1\chi_2+\chi_3\chi_1\right)
\end{equation}
are not in conflict with the VEV alignment, because they do not change the structure of the minimisation conditions at the minimum $\ev{\chi}=\left(v,\,v,\,v\right)$
\begin{equation}
0=\left[\frac{\partial V_\chi+V_{soft,Z_3}}{\partial\chi_i}\right]_{\chi_i=v^\prime}=\frac{2}{3} v^\prime \left(3 m_s^2+\sqrt{3} m_0^2+6 \lambda _1
   v^{\prime 2}\right)\;.
 \end{equation}
Hence, the flavon potential enforces the VEVs to align.

We thus conclude that the desired vacuum alignment requires a
mechanism to forbid all couplings between the flavon sectors that
break $A_4$ to $Z_2$ and $Z_3$, respectively, except for the quartic
coupling where both couple in pairs to singlets. This can be rephrased
in the requirement to have an 'accidental' symmetry $A_4^\nu\times A_4^f$ in
the flavon potential, where the first group factor, $A_4^\nu$, corresponds to the
flavons coupling to neutrinos and the second one, $A_4^f$, to flavons coupling
to charged leptons.
Note, that the Kronecker product $\Rep{3}\times
\Rep{3}=\MoreRep{1}{1}+\MoreRep{1}{2}+\MoreRep{1}{3}+\Rep{3}_{S}+\Rep{3}_{A}$
allows couplings of the form $(\chi \chi)_{\MoreRep{1}{2}}(\phi \phi)_{\MoreRep{1}{3}}$, $(\chi
\chi)_{\MoreRep{1}{3}}(\phi \phi)_{\MoreRep{1}{2}}$ and $(\chi \chi)_{\Rep{3}}(\phi \phi)_{\Rep{3}}$ in the
minimal $A_4$ model discussed above and the desired vacuum alignment is thus not possible.
These kind of couplings can not be forbidden by assigning $\chi$ or $\phi$ to a unitary representation of an additional internal symmetry group commuting with the flavour group, because $\chi^\dagger\chi$ and $\phi^\dagger\phi$ will always be invariant. In particular, it is not possible to solve it by introducing an additional commuting group factor, which is a discrete group or a compact Lie group.

For this reason, the VEV alignment problem has been mainly studied within the context of SUSY as well as brane constructions within extra-dimensional models circumventing this problem.
In the following section, we show how the required vacuum alignment can be achieved with an internal symmetry group by extending the flavour group in a non-trivial way. 


\section{Group Extensions and Vacuum Alignment}
\label{sec:GroupScan}

In the following, we explain the type of groups we are searching for
and why we are searching for these groups. We always use the
group $A_4$ as an example, but the arguments hold for any group. In a
first step, we directly extend the group by adding new
generators, which do not commute with the generators of the flavour
group. In the second subsection, we generalize our approach and
look for general group extensions. The group theoretical notions we use are defined in the footnotes of this section.

\subsection{Semidirect Product Groups}
To reproduce the success of $A_4$ models, we search for an extended flavour
group 
\begin{equation}
G=\ev{S,T,X_1,\dots,X_n\vert S^2=T^3=(ST)^3=r^X_\alpha(X_1,\,\dots,\,X_n)=r^\mathrm{mix}_\beta(S,\,T,\,X_1,\,\dots,\,X_n)=1}
\end{equation}
that contains $H=\ev{S,T\vert S^2=T^3=(ST)^3=1}\simeq A_4$ as a subgroup. $X_i$ denote the additional generators of the extended group and $r^X_\alpha$, $r^\mathrm{mix}_\beta$ with $\alpha=1,\dots,s_\alpha$ and $\beta=1,\dots,s_\beta$ additional relations. Note that there are no additional relations involving only $S$ and $T$. As we discussed in the last section, not all of the additional generators can commute with $H$. Therefore, there have to be relations $r^\mathrm{mix}_\beta$.
These generators will be needed to forbid the dangerous couplings discussed in the last section. Any discrete group that contains $H$ as subgroup can be written in this way. 

We further demand that there should be 
representations $\rho_{\textbf{i}}:G\rightarrow GL(V)$ with
\begin{align}
\rho_{\textbf{i}}(X_j)=\mathbbm{1} \quad \forall\,j=1,\dots,n
\end{align}
and $\rho_{\textbf{i}}(S)$ and $\rho_{\textbf{i}}(T)$ corresponding to the usual $A_4$ representations $\textbf{i}=\MoreRep{1}{1}$, $\MoreRep{1}{2}$, $\MoreRep{1}{3}$ and $\Rep{3}$, e.g.
\begin{align}
\rho_{\Rep{3}}(S)&=\left(\begin{array}{ccc}
1&0&0\\
0&-1&0\\
0&0&-1
\end{array}\right),& 
\rho_{\Rep{3}}(T)&=\left(\begin{array}{ccc}
0&1&0\\
0&0&1\\
1&0&0
\end{array}\right).
\end{align} 
If the SM fermions are assigned to these representations, the $A_4$ predictions for the mixing angles remain unchanged. The existence of the representation $\rho\equiv
\rho_{\Rep{3}}$ gives a first constraint on the flavour group $G$: 
The image of the representation $\rho$ is isomorphic to $H$, i.e.~$\mathrm{im}(\rho)\cong H$, and its kernel~\footnote{The
  \emph{kernel} of a representation $\rho$ is defined by $\ker\rho=\{g\in G
  \vert \rho(g)=\mathbbm{1}\}$.}  $N\equiv\ker \rho=\ev{X_1,...,X_n}$ is a
normal subgroup~\footnote{A \emph{normal} subgroup $N$ of a group
  $G$, denoted by $N\lhd G$, is a subgroup, which is invariant under
  conjugation by an arbitrary group element of $G$,
  i.e.~$gNg^{-1}=N$.} of $G$ with the quotient group~\footnote{The
  \emph{quotient group} $G/N$ is defined by the set of the left cosets
$gN$ with $g\in G$.} $G/N\cong\mathrm{im}(\rho)\cong H$ (by the first isomorphism theorem). $\rho$ thus essentially defines a surjective homomorphism~\footnote{A \emph{ (group) homomorphism} $\rho : G\rightarrow H$ is a mapping preserving the group structure, i.e.~$\rho(g_1 g_2)=\rho(g_1)\rho(g_2)\;\forall g_{1,2}\in G$. A \emph{surjective} homomorphism $\rho:G\rightarrow H$ has the additional property $\mathrm{im}(\rho)=H$. } from G onto H, which is the identity on H and whose kernel is N. 
Groups of this type are known as semidirect product groups~\footnote{A group $G$ is a \emph{semidirect product} of a subgroup $H$ and
  normal subgroup $N$ if there exists a homomorphism $G\rightarrow H$
  which is the identity on H and whose kernel is N. 
  The \emph{ direct product} $N\times H$ can be considered as a semidirect product, where $H$ is a normal subgroup of $G$ as well. The two factors $N$ and $H$
of a direct product commute.} $G=N\rtimes H$, which is a generalisation of a direct
product $N\times H$. One example of a semidirect product group is
$A_4\cong (Z_2\times Z_2)\rtimes Z_3$ itself. As $N$ and $H$ can not commute, $G$ can not be a direct product, $G\neq N\times H$. 
  
Once we have found such a group we can assign the lepton doublets, charged leptons and the flavon $\chi$ that couples to the charged lepton sector in the usual way to representations \Rep{3} and \MoreRep{1}{i}, $i=1,2,3$, while assigning the flavon $\phi$ of the neutrino sector to an irreducible representation of $G$, which is faithful~\footnote{A representation $\phi$ is \emph{ faithful}, if the homomorphism $\phi: G\rightarrow \mathrm{GL}(V)$ is injective. It is \emph{ faithful on a subgroup N}, if $\phi\vert_N$ is faithful.} on $N$, and contains $\Rep{3}$ in the Kronecker product $\phi^n$ at some order $n$. If this representation $\phi$ was not faithful on $N$, it would be possible to restrict to the smaller group $G/\ker\phi\vert_N$ (by the third isomorphism theorem), which leads to the same flavour structure, and study its predictions.
The problematic cross-couplings $(\chi \chi)_{\MoreRep{1}{2}}(\phi
\phi)_{\MoreRep{1}{3}}$, $(\chi \chi)_{\MoreRep{1}{3}}(\phi
\phi)_{\MoreRep{1}{2}}$ and $(\chi \chi)_{\Rep{3}}(\phi
\phi)_{\Rep{3}}$ can now be forbidden, provided that the Kronecker
product $\phi\times\phi$ does not contain the representations \Rep{3} as well
as \MoreRep{1}{2,3}. Thus, the flavon potential of $\phi$ and $\chi$
exhibits an 'accidental' symmetry $G\times H$ at the renormalizable
level. This accidental symmetry is broken to $G$ at
an higher order in the flavon potential. 

\begin{table}[bt]
\centering

\begin{tabular}{|c|c|c|c|c|}
\hline
Subgroup $H$ & Order of $G$ & \GAP &Structure Description & $Z(G)$\\ \hline
\multirow{7}{*}{$A_4$} 
&$ 96$ & $204$ & $Q_8 \rtimes A_4$ & $Z_2$\\\cline{2-5}
&$ 288$ & $860$ & $T^\prime \rtimes A_4$ & $Z_2$\\\cline{2-5}
&$ 384$ & $617$, $20123$ & $((Z_2 \times Q_8) \rtimes Z_2) \rtimes A_4$ & $Z_2$\\\cline{2-5}
&$ 576$ & $8273$ & $(Z_2 . S_4) \rtimes A_4$ & $Z_2$\\\cline{2-5}
&\multirow{2}{*}{$ 768$} & $1083945$ & $(Z_4 . Z_4^2) \rtimes A_4$ & $Z_4$\\
& & $1085279$ & $((Z_2 \times Q_{16})\rtimes Z_2) \rtimes A_4$ & $Z_2$\\\hline
\multirow{9}{*}{$S_4$} 
&$ 192$ & $1494$ & $Q_8 \rtimes S_4$ & $Z_2$\\\cline{2-5}
&\multirow{2}{*}{$ 384$} & $18133$, $20092$ & $(Z_2 \times Q_8) \rtimes S_4$ & $Z_2$\\
& & $20096$ & $((Z_4 \times Z_2) \rtimes Z_2) \rtimes S_4$ & $Z_4$\\\cline{2-5}
&\multirow{2}{*}{$ 576$} & $8282$ & $ T^\prime \rtimes S_4$ & $Z_2$\\
& & $8480$ & $(Z_3 \times Q_8) \rtimes S_4$ & $Z_6$\\\cline{2-5}
&$ 768$ & $1086052$, $1086053$ & $((Z_2 \times Q_8) \rtimes Z_2) \rtimes S_4$ & $Z_2$\\\cline{2-5}
&$ 960$ & $11114$ & $(Z_5 \times Q_8) \rtimes S_4$ & $Z_{10}$\\\hline
\multirow{4}{*}{$T^\prime$} 
&$ 192$ & $1022$ & $Q_8 \rtimes T^\prime$ & $Z_2^2$\\\cline{2-5}
&$ 648$ & $533$ & $\Delta(27) \rtimes T^\prime$ & $Z_3$\\\cline{2-5}
&$ 768$ & $1083573$, $1085187$ & $((Z_2 \times Q_8) \rtimes Z_2) \rtimes T^\prime$ & $Z_2^2$\\\hline
\end{tabular}
\caption{Candidate groups $G$ up to order $1000$ that may be written as non-trivial
  semidirect products $G=N\rtimes H$ for the groups
  $H=A_4$, $T_7$, $S_4$, $T^\prime$, $\Delta(27)$ and that lead to an enhanced symmetry in the scalar potential making the correct vacuum alignment possible. No such groups were found for $H=T_7$,  $\Delta(27)$. Details of the groups may be accessed using the computer algebra system \GAP~by using the command
  \texttt{SmallGroup(Order,GAP)}. $Q_8$ denotes the quaternion group,
  which is defined in \Secref{sec:SmallestGroup} and the generalized
  quaternion group of order $16$, $Q_{16}$, is defined by
  $Q_{16}=\braket{x,y\vert x^8=1, \,x^2=y^4,\,y^{-1}xy=y^{-1}}$. The
  expression of the form $N . H$ is the \GAP\ notation of a central
  extension, i.e.~$N$ is a normal subgroup of $G$, which is contained
  in the centre of $G$, and $H$ is the quotient group $G/N\cong
  H$. Note that there can be more than one semidirect product of $N$
  by $H$. }
\label{tab:candidates}
\end{table}

We thus systematically search for flavour groups $G$ containing a subgroup
$H$ and a normal subgroup $N$ satisfying $G/N\cong H (\cong A_4)$,
which lead to an 'accidental' symmetry $G\times H$ in the renormalizable part of the flavon potential.
Using the computer algebra system \GAP~\cite{GAP4:2011} and its \SmallGroups\ 
catalogue~\cite{SmallGroups:2011}, we have
performed a scan over all discrete groups $G$ up to order $1000$.
As the vacuum alignment problem is not specific to the group $A_4$, we have searched for
semidirect product groups $N\rtimes H$ with the desired properties for the groups $H=A_4$, $T_7$, $S_4$, $T^\prime$ and $\Delta(27)$, which are known to be interesting for flavour model building\footnote{All of these groups have size less than 30 and contain a three-dimensional  representation. There are two more groups with this property: $A_4\times Z_2$ and $Z_9 \rtimes Z_3$ \cite{Parattu:2010cy}. See also \cite{Grimus:2011ff} for a recent overview of finite groups, which are useful for flavour model building.}. We applied the following conditions:
\begin{enumerate}
\item $G=N\rtimes H\neq N\times H$ with $H$ being one of the groups $A_4$, $T_7$, $S_4$, $T^\prime$ or $\Delta(27)$;
\item there is an irreducible representation $\phi$, which is faithful on $N$;
\item $\phi^n$ contains \Rep{3} for some $n$;
\item there is an 'accidental' symmetry $G\times H$ in the renormalizable part of the flavon potential, i.e.~there are only couplings via the trivial singlet between $\chi$ and $\phi$ at the renormalizable level, e.g. only
  $(\chi^2)_{\MoreRep{1}{1}} (\phi^2)_{\MoreRep{1}{1}}$ exists for
  real representations $\chi$, $\phi$;
\end{enumerate}
It turns out that there are only candidates for $A_4$, $T^\prime$ or $S_4$ up to order 1000, which
are presented in \Tabref{tab:candidates}. Although, there are
semidirect product groups which fulfil the first three criteria for
$H=T_7$, or $H=\Delta(27)$, none of them leads to the desired
accidental symmetry in the scalar potential. This might be related to
the fact that these groups have complex three-dimensional
representations, and there are more couplings that would have to be
forbidden by the additional symmetries than in the case of $H=A_4$, $T^\prime$ and $S_4$, which have real three dimensional representations. Additionally, there are simply less groups up to order $1000$, which can be considered as an extension of $T_7$ or $\Delta(27)$ compared to the other groups. 

Looking at the list of candidate groups, we further note that the
normal subgroup $N$ is non-abelian for all our candidate groups. In addition, the defining homomorphism~\footnote{Equivalently to the previous definition, a semidirect product $N\rtimes H$ can be defined via a homomorphism $\varphi : H \rightarrow \mathrm{Aut}(N)$, where Aut($N$) denotes the group of all \emph{automorphisms} of $N$, i.e. the isomorphisms $N\rightarrow N$. The \emph{defining homomorphism} is sometimes  indicated as index of $\rtimes$, i.e. $N\rtimes_\varphi H$.} of each semidirect product is injective for $H=A_4,\,S_4$~\footnote{The same applies for the wreath product $S_3^4\rtimes A_4$ introduced by Babu and Gabriel~\cite{Babu:2010bx}.} and in case of $H=T^\prime$, each group $N\rtimes T^\prime$ allows for a defining homomorphism with image $A_4$ or $T^\prime$.
The quaternion group $Q_8$, which frequently appears in \Tabref{tab:candidates}, is the smallest non-abelian group allowing for a defining homomorphism with these properties. Furthermore, all candidate groups have a non-trivial centre~\footnote{The \emph{centre} of a group, $Z(G)$, is
  the set of elements, which commute with all elements of the group $G$,
  i.e.~\mbox{$Z(G)\equiv\{x\in G\,\vert\, g x = x g\quad \forall g \in G\}$}. It
  forms a normal subgroup of $G$, i.e.~$Z(G)\lhd G$.}  $Z(G)\lhd N$. Hence, the representations can be classified according to their way of representing the elements in the centre, i.e.~whether (a subgroup of) the centre is represented trivially (mapped to the identity) or not. In particular, the representations $\chi$ of $G$, which are directly related to irreducible representations $\chi^H$ of $H$ with $\chi\vert_H\equiv\chi^H$ map the centre to the identity. They are single valued (in analogy to the representations of $\SU{2}$ with integer spin).
However, groups that fulfil these conditions do not necessarily have to have a non-trivial centre. For example the wreath product $S_3^4\rtimes A_4$, introduced by Babu and Gabriel~\cite{Babu:2010bx}, has a trivial centre.

Before studying the vacuum alignment for the smallest candidate group, let us look more closely at
how the breaking to different subgroups leads to the flavour
structure. As has been mentioned in the introduction, it has been
argued in \cDifferentSubgroups\ that the neutrino mixing matrix can be obtained by breaking the flavour
group to different subgroups in the charged and neutral fermion
sector, respectively\footnote{The role of the unbroken subgroups in neutrino mixing has also been discussed from a bottom-up perspective in \cite{Lam:2008sh}.}. It is usually broken by flavoured scalar fields
acquiring a VEV, which breaks the flavour group to the corresponding
little group\footnote{The \emph{little group} $G_{\ev{\phi}}$ is the subgroup of $G$ leaving a VEV $\ev{\phi}$ invariant, i.e.~$G_{\ev{\phi}}=\left\{g\in G\vert g\ev{\phi}=\ev{\phi}\right\}$. It is also denoted by \emph{stabilizer subgroup} or \emph{isotropy group}.}. However, the little group of the VEV of a scalar field
is not necessarily the little group relevant for the flavour
structure, as the mass term might be generated by a Kronecker product
of several scalar fields, e.g. the neutrino mass matrix might be given
by $\ell H \ell H \braket{\phi}^2$ and, therefore, the little group of
$\braket{\phi}^2$ is the relevant one~\footnote{For model building of this type, see \cite{King:2009lr}.}.
More concretely, in the case of $Q_8\rtimes A_4$, the existence of the non-trivial centre implies that neutrino masses are generated via $(\ell H
\ell H)\phi^{2n}$ for some $n>0$ and the little group of $\braket{\phi}^{2n}$ is enlarged by the centre to $\braket{S,\,Z(Q_8\rtimes A_4)}$. A similar reasoning applies to the other candidate groups.
Note that, so far we only investigated one flavon $\phi$ in an irreducible representation, which does not apply in the more general discussion with multiple flavons $\phi_i$ (or equivalently a reducible representation $\phi$). In this more general setup, the relevant combination of flavon VEVs contributing to the neutrino mass matrix can break the invariance again. Ultimately, the minimisation of the flavon potential decides which VEV alignment is achieved.

\subsection{General Group Extensions}

Let us have a closer look at the construction in the last section. In order to obtain the same flavour structure within $G$ as within $H$, we demanded the existence of representations $\rho_{\textbf{i}}$, which are directly related to the representations $\rho^H_{\textbf{i}}$ of $H$. The representations $\rho_{\textbf{i}}$ can be explicitly constructed using the surjective homomorphism from $G$ to $H$, which we will denote by $\xi:G\rightarrow H$:
\begin{equation*}
\rho_{\textbf{i}}\equiv\rho^H_{\textbf{i}}\circ \xi\;.
\end{equation*}
Hence, as soon as there is a surjective homomorphism $\xi:G\rightarrow
H$, there are representations $\rho_{\textbf{i}}$ with the desired
property. Therefore, is it enough to look for groups $G$ and a surjective homomorphism $\xi:G\rightarrow H$. This automatically implies the existence of a normal subgroup $N=\ker\xi$ and a quotient group $G/N\cong H$. Thus, we are only dropping the condition that $H$ is a subgroup of $G$.
Actually, this type of extension is a general problem in group theory, which aims to
find all possible groups $G$ given two groups $N$ and $H$, such that
$G/N\cong H$. In the mathematical literature, this is denoted by \emph{short exact sequence}.
One example of such an extension is $T^\prime$.
$A_4$ is not a subgroup of $T^\prime$, but $A_4\cong T^\prime/Z_2$. 
In $T^\prime$ models~\cTprime, the flavour structure of the lepton sector is
essentially described by the quotient group $T^\prime/Z_2\cong A_4$ and the additional
group structure, i.e.~the two dimensional representations
\MoreRep{2}{i}, are used to describe the quark sector. Hence, group extensions of the kind we described are not limited to the VEV alignment, but can be used more generally to lift properties of one group $H$ to a larger group $G$, which addresses additional questions in flavour physics. Therefore, we propose to use these kind of constructions more systematically.

\begin{table}[bt]
\centering
\begin{tabular}{|c|c|c|c|}
\hline
Quotient Group $H$ & Order of $G$ & \GAP &Structure Description\\ \hline
\multirow{3}{*}{$A_4$}
&$96$ & $201$ & $Z_2 . (Z_2^2 \times A_4)  $\\\cline{2-4}
&$144$ & $127$ & $Z_2 . (A_4 \times S_3)  $\\\cline{2-4}
&$192$ & $1017$ & $Z_2 . (D_8 \times A_4)  $\\\hline
\multirow{6}{*}{$S_4$}
&$96$ & $67$, $192$ & $Z_4  .  S_4 $\\\cline{2-4}
&$144$ & $121$, $122$ & $Z_6  .  S_4  $\\\cline{2-4}
&$192$ & $187$, $963$ & $Z_8  .  S_4  $\\\cline{2-4}
&$192$ & $987$, $988$ & $Z_2 . ((Z_2^2 \times A_4) \rtimes Z_2)  $\\\cline{2-4}
&$192$ & $1483$,$1484$ & $Z_2 . (Z_2^2 \times S_4)  $\\\cline{2-4}
&$192$ & $1492$ & $Z_2 . ((Z_2^4 \rtimes Z_3) \rtimes Z_2)  $\\\hline
$T^\prime$&$192$ & $1007$ & $Z_2^2  . (Z_2^2 \times A_4)  $\\\hline
\end{tabular}
\caption{Candidate groups $G$ up to order $200$, which can not be written as semidirect product. The expression of the form $N . H$ in the last column is the \GAP\ notation of a central extension, i.e.~$N$ is a normal subgroup of $G$, which is contained in the centre of $G$, and $H$ is the quotient group $G/N\cong H$. Here, we explicitly choose $N=Z(G)$ and therefore $N . H = Z(G) . G/Z(G)$. The candidate groups of order 200-500 can be found in \Tabref{tab:candGeneralRest}.}
\label{tab:candidatesGeneral}
\end{table}

In this article, however, we are mainly interested in a solution to
the vacuum alignment problem, and therefore, we do not consider these
other possibilities further, but perform another scan looking for
groups solving the vacuum alignment problem and we relaxed the first condition of the previous scan to
\begin{enumerate}
\item $G/N\cong H$ with $H$ being one of the groups $A_4$, $T_7$, $T^\prime$~\footnote{We included $T^\prime$ in this scan, although $T^\prime$ is an extension of $A_4$ via $T^\prime/Z_2\cong A_4$. However, the second condition excludes several candidates for $T^\prime$, because the $Z_2$ in $T^\prime/Z_2\cong A_4$ is a subgroup of the $N$ in the second condition.}, $S_4$, $\Delta(27)$,
\end{enumerate}
while keeping the other conditions. It turns out that there are only
candidates for $A_4$, $T^\prime$ and $S_4$ up to order $1000$. We
collect all candidates up to order $200$, which are not contained in
the previous search for semidirect product groups, in
\Tabref{tab:candidatesGeneral} and present the candidates of order $200-500$ in \Tabref{tab:candGeneralRest}. 


\section{Smallest Group: $Q_8\rtimes A_4$}
In this section, we discuss a model of lepton masses and mixings based on the smallest semidirect product group $Q_8\rtimes A_4$ in the catalogue obtained in the preceding section. After a brief description of the group, we discuss why it is necessary to employ more than one faithful representation of the full group and build a model with this particle content. We then show that the most general scalar potential has the desired accidental symmetry and thus allows for the correct vacuum alignment.
\label{sec:SmallestGroup}
\subsection{Group Theory}
While the $A_4$ subgroup is presented by
\begin{align}
\braket{S,T\vert S^2=T^3=(ST)^3=1}\;,
\end{align}
the quaternionic subgroup $Q_8$ (also known as $D^\prime_4$, the double group of the dihedral group of order 4) is defined by 
\begin{align}
\braket{X,Y\vert X^4=1,\; X^2=Y^2,\; Y^{-1}XY=X^{-1}}\;. 
\end{align}
The semidirect product $Q_8\rtimes A_4$ we are considering here is defined by the additional relations between the generators of $Q_8$ ($X$, $Y$) and $A_4$ ($S$, $T$)
\begin{align}
SXS^{-1}&=X, & SYS^{-1}&=Y^{-1}, & 
TXT^{-1}&=YX, & TYT^{-1}&=X\;.
\end{align}

An explicit matrix representation  of these generators  for the relevant representations is given in Table
\ref{tab:Q8rtimesA4representations} and the character table is presented in Table \ref{tab:ctblQ8A4}. The Kronecker products
\begin{subequations}
\begin{align}
\MoreRep{3}{i}\times\MoreRep{3}{i}&=\MoreRep{1}{1}+\MoreRep{1}{2}+\MoreRep{1}{3}+{\MoreRep{3}{i}}_S+{\MoreRep{3}{i}}_A\\
\MoreRep{3}{i}\times\MoreRep{3}{j}&=\sum_{\stackrel{k=1}{k\neq i,j}}^{5}\MoreRep{3}{k} \hspace{2cm} (i\neq j)\\
\MoreRep{3}{i}\times\MoreRep{4}{j}&=\MoreRep{4}{1}+\MoreRep{4}{2}+\MoreRep{4}{3}\\
\MoreRep{4}{1}\times\MoreRep{4}{1}&={\MoreRep{1}{1}}_S+{\MoreRep{3}{1}}_A+{\MoreRep{3}{2}}_S+{\MoreRep{3}{3}}_S+{\MoreRep{3}{4}}_S+{\MoreRep{3}{5}}_A\\
\MoreRep{4}{1}\times\MoreRep{4}{2}&={\MoreRep{1}{2}}_S+{\MoreRep{3}{1}}_A+{\MoreRep{3}{2}}_S+{\MoreRep{3}{3}}_S+{\MoreRep{3}{4}}_S+{\MoreRep{3}{5}}_A
\end{align}
\end{subequations}
show that if one uses the unfaithful triplet $\chi\sim \MoreRep{3}{1}$ to break $A_4$ in the charged lepton sector and the four dimensional faithful representation $\phi\sim \MoreRep{4}{1}$ in the neutrino sector, there are no dangerous cross-coupling terms of the form $(\phi \phi)_{\MoreRep{3}{1}}(\chi \chi)_{\MoreRep{3}{1}}$ etc. allowed by the symmetry that would forbid the required VEV alignment.

\begin{table}
\centering
\begin{tabular}{|c|c|c|c|c|}
\hline  & S & T&X&Y \\ 
\hline $\MoreRep{1}{1}$ & $1$ & $1$&$1$ & $1$   \\ 
\hline $\MoreRep{1}{2}$ & $1$ & $\omega$&$1$ &$1$    \\ 
\hline $\MoreRep{1}{3}$ & $1$ & $\omega^2$ &$1$ & $1$  \\ 
\hline $\MoreRep{3}{1}$ & $\left(\begin{array}{ccc}
1&0&0\\
0&-1&0\\
0&0&-1
\end{array}\right)$ & $\left(\begin{array}{ccc}
0&1&0\\
0&0&1\\
1&0&0
\end{array}\right)$ & $\left(\begin{array}{ccc}
1&0&0\\
0&1&0\\
0&0&1
\end{array}\right)$&$\left(\begin{array}{ccc}
1&0&0\\
0&1&0\\
0&0&1
\end{array}\right)$\\
\hline \hline
$\MoreRep{4}{1}$&$\left(\begin{array}{cccc}
0&1&0&0\\
1&0&0&0\\
0&0&0&-1\\
0&0&-1&0
\end{array}\right)$&$\left(\begin{array}{cccc}
0&1&0&0\\
0&0&1&0\\
1&0&0&0\\
0&0&0&1
\end{array}\right) $&
$\left(\begin{array}{cccc}
0&0&-1&0\\
0&0&0&1\\
1&0&0&0\\
0&-1&0&0
\end{array}\right)$&$\left(\begin{array}{cccc}
0&1&0&0\\
-1&0&0&0\\
0&0&0&1\\
0&0&-1&0\end{array}\right)$\\ \hline
\end{tabular} 
\caption{Relevant Representations of $Q_8\rtimes A_4$ in some basis. The first 4 representations are the unfaithful $A_4$ representations the leptons are assigned to (therefore $X=Y=\mathbbm{1}$). The last representation is used to break $A_4$ in the neutrino sector. Note that this representation is double valued, i.e. $X^2=Y^2=-\mathbbm{1}$. Here $\omega=e^{\I 2 \pi/3}$.\label{tab:Q8rtimesA4representations}}
\end{table}

The relevant operators for the generation of the lepton masses are $\ell \chi
f^c\tilde{H}$ with $f$ being $e$, $\mu$ or $\tau$ as well as $\ell H \ell H$ and $\ell
H\ell H\phi^4$ for neutrino masses~\footnote{Here we again assume the discrete $Z_2$ symmetry
$\chi\rightarrow-\chi$, $f^c\rightarrow-f^c$ to separate the charged from the
neutral fermion sector.}.

Unfortunately, the most general VEV configurations $\phi\sim(a,a,b,-b)$ that break the group to the $Z_2$ subgroup generated by $S$ cannot be realised in the flavon potential\footnote{The operator $(\phi \phi) _{\MoreRep{3}{4}} \cdot (\phi \phi) _{\MoreRep{3}{4}}$, which one would naively expect, can be expressed as a linear combination of the other operators.} 
\begin{align}
V_\phi(\phi)&=\mu_1^2 (\phi \phi)_{\MoreRep{1}{1}}+\alpha_1 (\phi \phi)_{\MoreRep{1}{1}}^2 + \sum_{i=2,3}\alpha_i (\phi \phi) _{\MoreRep{3}{i}} \cdot (\phi \phi) _{\MoreRep{3}{i}}
\end{align}
due to the relation 
\begin{align}
0=b \left.\frac{\partial V_\phi}{\partial \phi_1}\right|_{\braket{\phi}}-a\left.\frac{\partial V_\phi}{\partial \phi_3}\right|_{\braket{\phi}}=\frac{4}{\sqrt{3}}a b (a^2-b^2)(\alpha_2+\alpha_3)\;.
\end{align}
The achievable VEV configurations with $a^2=b^2$ or $ab=0$ lead to a restoration
of symmetry in the operator $(\ell \ell)_{\MoreRep{3}{1}}\left( \phi^4
\right)_{\MoreRep{3}{1}}$ that generates the $(\ell \ell)_{\MoreRep{3}{1}}$ entry in the mass matrix and consequently it vanishes in the vacuum, $\langle\left( \phi^4 \right)_{\MoreRep{3}{1}}\rangle\sim a b (a^2-b^2)$\footnote{If one introduces a soft-breaking term that conserves the $Z_2$ subgroup generated by S, $V_S=\alpha \left( \phi_1\phi_2+  \phi_3\phi_4  \right)$ in the potential, the minimum with $a\neq b$ can then be realised. We do not pursue this option further here, as we are interested in genuine spontaneous symmetry breaking.}.

This type of model is also not so interesting from a general point of view, as it shares a couple of unpleasant features with the model of Babu and Gabriel\cite{Babu:2010bx} when viewed as an effective field theory:
\begin{itemize}
\item the off-diagonal entries in the neutrino mass matrix, generated by $(\ell H \ell H \phi^4)$, would be of very different order than the diagonal ones generated by the operator $(\ell H \ell H)$. To satisfy neutrino data, the two entries have to be of almost the same size, though.
\item as $(\ell H \ell H \chi^2)$ is allowed and of smaller dimension than $(\ell H \ell H \phi^4)$, tri-bimaximal mixing is not a leading-order prediction of the model.
\end{itemize}
All of these issues can of course be cured by introducing a UV completion that does not confirm the effective field theory prejudices.

However, we restrict ourselves to natural solutions within effective field theory. To solve all of these problems, we will discuss a model with two flavons in the neutrino sector, $\phi_1\sim \MoreRep{4}{1}$ and $\phi_2\sim\MoreRep{4}{1}$, where an additional symmetry forbids the allowed term $\chi \cdot (\phi_1 \phi_2)_{\MoreRep{3}{1}}$ that could disturb the VEV alignment between the two sectors. We identify this symmetry with the one that separates the charged lepton from the neutral lepton sector, i.e. we postulate the additional $Z_4$ symmetry $\ell\rightarrow \I \ell$, $f^c \rightarrow -\I f^c$ and $\phi_2\rightarrow -\phi_2$, where $f$ denotes $e$, $\mu$ and $\tau$. One can think of this $Z_4$ symmetry as a discrete version of lepton number with $\phi_2$ being doubly charged under this lepton number.

\begin{table}
\centering
\begin{tabular}{c|ccccccccccc}
&$1$&$T$&$ SYX$ &$SY$&$Y^2$&$T^2$&$TY$&$S$&$SX$&$X$&$STYT$ \\ \hline
$\MoreRep{1}{1}    $&  1&  1 & 1&  1 & 1 & 1 &  1 & 1  &1 & 1  & 1  \\
$\MoreRep{1}{2}    $&  1& $\omega$&  1&  1&  1&  $\omega^2$&  $\omega$&  1&  1&  1&   $\omega^2$ \\
$\MoreRep{1}{3}    $&  1&  $\omega^2$ & 1&  1 & 1 &$\omega$&   $\omega^2$&  1 & 1&  1 & $\omega$ \\
$\MoreRep{3}{1}$&      3&  . &-1& -1&  3&  . &  . &-1 &-1 & 3  & . \\
$\MoreRep{3}{2}$&      3&  . & 3 &-1 & 3 & .  & . &-1& -1 &-1  & . \\
$\MoreRep{3}{3}$&      3&  . &-1&  3 & 3 & .  & . &-1& -1 &-1  & . \\
$\MoreRep{3}{4}$&      3&  . &-1& -1&  3&  . &  .&  3& -1& -1  & . \\
$\MoreRep{3}{5}$&      3&  . &-1& -1&  3&  . &  .& -1&  3& -1  & . \\
$\MoreRep{4}{1}$&      4&  1&  . & . &-4&  1&  -1&  .&  .&  .&  -1 \\
$\MoreRep{4}{2}$&     4 & $\omega^2$ & .  &. &-4 &$\omega$&  -$\omega^2$&  .&  .&  .& -$\omega$ \\
$\MoreRep{4}{3}$&     4 &$\omega$&  . & . &-4&  $\omega^2$& -$\omega$&  .&  .&  .&  -$\omega^2$ \\
\end{tabular}
\caption{Character table of $Q_8\rtimes A_4$. The first line are
  representatives of the different conjugacy classes. Zeroes in the
  character table are denoted by a dot $.$  and $\omega$ is the third
  root of unity $\omega=e^{2\pi\I/3}$.\label{tab:ctblQ8A4}}
\end{table}

\subsection{Model and Lepton Masses}
Finally, we present a model based on the symmetry group $Q_8\rtimes A_4$ augmented by the auxiliary symmetry $ Z_4$ introduced at the end of the last section. The
leptonic and scalar particle content is given in
\Tabref{tab:partcontent}. As advertised, for the standard model leptons, we use the unfaithful representations $\MoreRep{1}{1,2,3}$ and $\MoreRep{3}{1}$, that transform as irreducible representations under the subgroup $A_4$. In the charged sector we use the unfaithful representation $\chi\sim \MoreRep{3}{1}$ and the charged lepton sector is thus
analogous to the usual construction in an $A_4$ model. In the neutrino sector, we introduce the real flavons $\phi_{1,2}\sim \MoreRep{4}{1}$.

 To keep the discussion simple, we use an effective field theory description. To lowest order, the charged lepton masses arise from the operators
\begin{equation}
\mathcal{L}_e^{(5)} = y_e (\ell \chi)_{\MoreRep{1}{1}} e^c \tilde{H}/\Lambda
+y_\mu (\ell \chi)_{\MoreRep{1}{3}} \mu^c  \tilde{H}/\Lambda
+y_\tau (\ell \chi)_{\MoreRep{1}{2}} \tau^c  \tilde{H}/\Lambda +\hc\; ,
\label{eq:chargedlepton}
\end{equation}
with $\tilde{H}=\I \sigma_2 H^*$, and the neutrino masses are generated from the effective interactions
\begin{align}
\mathcal{L}_\nu^{(7)} &= x_{a} (\ell H\ell H)_{\MoreRep{1}{1}} (\phi_1 \phi_2)_{\MoreRep{1}{1}} /\Lambda^3 + x_d (\ell H\ell H)_{\MoreRep{3}{1}} \cdot(\phi_1 \phi_2)_{\MoreRep{3}{1}} /\Lambda^3 +\hc\;. \label{eq:neutrinomassop}
\end{align}
The notation should be self-explanatory and the relevant Kronecker products are given in appendix \ref{app:GTofQ8}.
We will show in the next section that the vacuum configuration 
\begin{align}
\langle\chi\rangle &=(v',v',v')^T,&
\langle\phi_1\rangle &=\frac{1}{\sqrt{2}}(a,a,b,-b)^T,&
\langle\phi_2\rangle &=\frac{1}{\sqrt{2}}(c,c,d,-d)^T
\end{align}
with $v',a,b,c,d \in \mathbb{R}$, can be obtained as the global minimum of the most general scalar potential. This configuration gives $\langle  (\phi_1 \phi_2) _{\MoreRep{3}{1}} \rangle=\frac12(b c-a d,0,0)^T$ and $\langle  (\phi_1 \phi_2) _{\MoreRep{1}{1}} \rangle=\frac12(a c+b d)$ and it is of course non-unique as there are many more physically identical patterns that can be obtained by acting with the group generators on this vacuum. This VEV configuration breaks the flavour symmetry to the $Z_2$ subgroup generated by $S$. There are also physically inequivalent minima of the potential that break to the $Z_2$ subgroups generated by $SY$ and $SYX$ which lead to the same structure $\langle  (\phi_1 \phi_2) _{\MoreRep{3}{1}} \rangle\propto (1,0,0)^T$. We will comment on this more in \Secref{sec:scalar potential}. 

The leading-order mass matrices are given by 
\begin{align}
M_E=\frac{v' v}{\sqrt{2}\Lambda }\left( \begin{array}{ccc}
y_e&y_\mu & y_\tau\\
y_e &\omega y_\mu &\omega^2 y_\tau\\
y_e &\omega^2 y_\mu &\omega y_\tau\\
\end{array}\right), \hspace*{1cm}
m_\nu=\frac{v^2 }{2\sqrt{3} \Lambda^3 }\left( \begin{array}{ccc}
\tilde{a} & 0 & 0\\
0&\tilde{a} & \tilde{d} \\
0 &\tilde{d} &\tilde{a} \\
\end{array}\right), \hspace*{1cm}
\label{eq:numass}
\end{align}
with 
\begin{align}
\tilde{a} & =  x_{a} \frac12(a c+b d) , & \tilde{d} &= x_{d} \frac12
(b c-a d)&  &\mathrm{and}&
 \langle H\rangle &=\left(\begin{array}{c} 0\\ v/\sqrt{2}
\end{array}
\right). 
\end{align}

The mass matrices can be diagonalized by 
\begin{align}
U_0&=\frac{1}{\sqrt{3}}
\left(
\begin{array}{ccc}
 1 & 1 & 1 \\
 1 & \omega  & \omega^2  \\
 1 & \omega ^2 & \omega 
\end{array}
\right),
& 
U_\nu
&=\left(
\begin{array}{ccc}
 0 & 1 & 0 \\
 \frac{1}{\sqrt{2}} & 0 & -\frac{i}{\sqrt{2}} \\
 \frac{1}{\sqrt{2}} & 0 & \frac{i}{\sqrt{2}}
\end{array}
\right)
\end{align}
such that $U_0^{\dagger} M_e =\frac{v' v}{\sqrt{2}\Lambda} \diag(y_e,y_\mu,y_\tau)$ \footnote{The charged lepton mass hierarchy can be explained by a Froggatt-Nielsen $\U{1}$ symmetry in the usual way.}, $U_\nu^T M_\nu U_\nu=\frac{v^2}{2\sqrt{3} \Lambda^3 }\diag(\tilde{a}+\tilde{d},\tilde{a},-\tilde{a}+\tilde{d})$ and the resulting mixing matrix $U_{MNS}=U_0^\dagger U_\nu$ is given by the HPS matrix (\ref{eq:HPSmatrix}). This construction is of course completely analogous to the usual $A_4$ models\cite{Altarelli:2005yx} and it is well-known that the moderate tuning $\vert \tilde{a}\vert\sim \vert \tilde{d}\vert$ is needed in order to accommodate the correct neutrino spectrum\cite{Brahmachari:2008lr,*Barry:2010fk}. 
However, in the usual $A_4$ models the contributions to $\tilde{a}$ and $\tilde{d}$ stem from completely different VEVs, while in our model both stem from VEVs of the same fields, and a similar order of magnitude might therefore be considered more natural. Indeed, in the numerical minimisation of the potential, we found a tendency for a similar size of the two $\phi$ contractions. 

Let us also comment on the fact that we have to employ two real copies of the faithful representation $\MoreRep{4}{1}$.
This exactly matches the numbers of degrees of freedom of one complex $A_4$ triplet and one complex singlet, which is commonly used (e.g. \cite{Altarelli:2005yx}). Here, we do not have to introduce additional degrees of freedom to obtain the correct vacuum alignment and we thus think it is an attractive and economical model. 

The effects of higher order operators can be found in \Secref{sec:higherorderop}.

\begin{table}
\begin{center}
\begin{tabular}{|l|ccc||c|c|}\hline
particle & ${SU}({3})_c$ & ${SU}({2})_L$ & ${U}({1})_Y$&
$Q_8\rtimes A_4$&$Z_{4}$\\\hline\hline
$\ell$ & 1 & 2 & -1/2  &\MoreRep{3}{1}&$\I$\\
$e^c+\mu^c+\tau^c$ & 1 & 1 & 1   & $\MoreRep{1}{1}+\MoreRep{1}{2}+\MoreRep{1}{3}$&$-\I$ \\\hline
$H$ & 1 & 2 & 1/2  &\MoreRep{1}{1} &1 \\\hline
$\chi$ & 1 & 1 & 0  &\MoreRep{3}{1} &1 \\\hline
$\phi_1$ & 1 & 1 & 0  &\MoreRep{4}{1} &1\\
$\phi_2$ & 1 & 1 & 0  &\MoreRep{4}{1}&$-1$ \\\hline
\end{tabular}
\caption{Particle content of the minimal model with the correct spontaneous symmetry breaking. All fermions are left-handed Weyl fermions.  \label{tab:partcontent}}
\end{center}
\end{table}

\subsection{Scalar Potential}
\label{sec:scalar potential}
Here, we demonstrate that the pattern of vacuum expectation values we used in the last section can be obtained as the minimum of the scalar potential. The most general scalar potential invariant under the flavour symmetry is given by
\begin{align}
V(\chi,\phi_1,\phi_2)=V_\chi(\chi)+V_\phi(\phi_1,\phi_2)+V_{\mathrm{mix}}(\chi,\phi_1,\phi_2)
\end{align}
with
\begin{align}
V_\phi(\phi_1,\phi_2)=&\mu_1^2 (\phi_1 \phi_1)_{\MoreRep{1}{1}}+\alpha_1 (\phi_1 \phi_1)_{\MoreRep{1}{1}}^2 + \sum_{i=2,3}\alpha_i (\phi_1 \phi_1) _{\MoreRep{3}{i}} \cdot (\phi_1 \phi_1) _{\MoreRep{3}{i}}\nonumber \\
+&\mu_2^2 (\phi_2 \phi_2)_{\MoreRep{1}{1}}+\beta_1 (\phi_2 \phi_2)_{\MoreRep{1}{1}}^2 + \sum_{i=2,3}\beta_i (\phi_2 \phi_2) _{\MoreRep{3}{i}} \cdot (\phi_2 \phi_2) _{\MoreRep{3}{i}}\nonumber\\
+&\gamma_1 (\phi_1 \phi_1)_{\MoreRep{1}{1}} (\phi_2 \phi_2)_{\MoreRep{1}{1}}+ \sum_{i=2,3,4}\gamma_i (\phi_1 \phi_1) _{\MoreRep{3}{i}} \cdot (\phi_2 \phi_2) _{\MoreRep{3}{i}}\nonumber\\
V_\chi(\chi)&=\mu^2_3 (\chi \chi)_{\MoreRep{1}{1}}+\rho_1 (\chi \chi \chi)_{\MoreRep{1}{1}}+ \lambda_1 (\chi \chi)_{\MoreRep{1}{1}}^2+\lambda_2  (\chi \chi)_{\MoreRep{1}{2}}(\chi \chi)_{\MoreRep{1}{3}}
\nonumber\\
V_{\mathrm{mix}}(\chi,\phi_1,\phi_2)&=\zeta_{13} (\phi_1 \phi_1)_{\MoreRep{1}{1}} (\chi \chi)_{\MoreRep{1}{1}} +\zeta_{23}(\phi_2 \phi_2)_{\MoreRep{1}{1}}  (\chi \chi)_{\MoreRep{1}{1}}
\end{align}
Note that, by construction, there are no non-trivial couplings between the $\chi$ and $\phi$ breaking sectors that would disturb the vacuum alignment. The potential thus has an 'accidental' $[(Q_8\rtimes A_4)\times A_4]\times Z_4$ symmetry. This symmetry is explicitly broken to $(Q_8\rtimes A_4)\times Z_4$ by the couplings to leptons and by higher dimensional operators in the potential. As the accidental symmetry is discrete, there is no pseudo-Goldstone boson, as can easily happen in constructions of this type~\cite{Babu:2010bx}.

Let us now demonstrate that this model does not suffer from a vacuum alignment problem. At first, we discuss the possible minima of the potential focussing on the little group in the neutrino sector, i.e. the subgroup, which leaves the VEV invariant. If there is a minimum, in which the symmetry generator $Q \in G$ is  left unbroken, i.e. $Q\ev{\phi_{1,2}}=\ev{\phi_{1,2}}$,  there obviously are degenerate minima $\ev{\tilde{\phi}_{1,2}}=g \ev{\phi_{1,2}}$ that leave $g Q g^{-1}$ unbroken, with $g \in G$. The physically distinct minima are therefore characterised by the conjugacy class(es) $G\cdot Q_i=\{ gQ_ig^{-1}\vert g \in G \} $ of the group element(s) $Q_i$.  Obviously, only conjugacy classes with an eigenvalue $+1$ can lead to a non-trivial little group.
For the four dimensional representation \MoreRep{4}{1}, there are five such classes which are represented by $1$, $S$, $SY$, $SYX$, $T$ as well as $T^2$. The groups generated by $T$ and $T^2$ are identical. For the three dimensional representation \MoreRep{3}{1}, where $X$ and $Y$ are represented trivially, all conjugacy classes have an eigenvalue $+1$ and can lead to a non-trivial little group. The relevant little group in the neutrino sector is the one of $\ev{\left(\phi_1\phi_2\right)_{\MoreRep{3}{1}}}$.

In the following, we will firstly discuss the possible little groups of $\ev{\phi_i}$ and then its implications for the little group of $\ev{\left(\phi_1\phi_2\right)_{\MoreRep{3}{1}}}$.
There are three physically distinct minima of $\phi_1$, that preserve a $Z_2$ subgroup:
\begin{itemize}
 \item $\ev{\phi_1}=\frac{1}{\sqrt{2}}(a,a,b,-b)^T$ results in the little group $\ev{S}$,
 \item  $\ev{\phi_1}=(0,a,b,0)^T$ in $\ev{SY}$ and
 \item  $\ev{\phi_1}=\frac{1}{\sqrt{2}}(-a,b,-a,b)^T$ in $\ev{SYX}$\;.
\end{itemize}
In addition, there is one preserving a $Z_3$ subgroup:
\begin{itemize}
 \item $\ev{\phi_1}=\frac{1}{\sqrt{2}}(a,a,a,b)^T$ preserves $\ev{T}$ (as well as $\ev{T^2}=\ev{T}$).
\end{itemize}
Obviously, there are also minima leading to little groups, which are generated by more than one generator. For example $\ev{\phi_1}\propto(1,1,1,-1)^T$ preserves $\ev{S,\,T}\cong A_4$. The same discussion applies to $\phi_2$.
The little group of $\ev{\left(\phi_1\phi_2\right)_{3_1}}$ contains the intersection of the little groups of $\ev{\phi_1}$ and $\ev{\phi_2}$. 

In the following, we will concentrate on the three little groups $\ev{S}$, $\ev{SY}$ and $\ev{SYX}$, which we listed above. If both $\ev{\phi_1}$ and $\ev{\phi_2}$ preserve the same $Z_2$ subgroup, we obtain $\langle  (\phi_1 \phi_2) _{\MoreRep{3}{1}} \rangle=\frac12(b c-a d,0,0)^T$ and $\langle  (\phi_1 \phi_2) _{\MoreRep{1}{1}} \rangle=\frac12(a c+b d)$ due to $X=Y=\mathbbm{1}$ for the SM representations (\MoreRep{1}{i} and \MoreRep{3}{1}) with $a, b$ being the VEVs of $\phi_1$ and $c,d$, the corresponding ones of $\phi_2$. Thus, it is impossible to distinguish these minima from low-energy neutrino phenomenology at the leading order. They are, however, physically distinct, since $\ev{\phi_1\phi_1}$ as well as $\ev{\phi_2\phi_2}$ are different and they lead to different mass spectra in the scalar sector.
The minimisation conditions for the VEVs are:
\begin{subequations}
\begin{align}
a\left( \alpha _+ \left(a^2+b^2\right)+\alpha _- \left(a^2-b^2\right)+\gamma _+ \left(c^2+d^2\right)+\gamma _- \left(c^2-d^2\right)+U_1\right) +\Gamma b c d &=0\\
b\left(\alpha _+ \left(a^2+b^2\right)- \alpha _- \left(a^2-b^2\right)+\gamma _+ \left(c^2+d^2\right)-\gamma _- \left(c^2-d^2\right)+U_1\right) +\Gamma a c d&=0 \\
c\left(\beta _+ \left(c^2+d^2\right)+\beta _- \left(c^2-d^2\right)+\gamma _+ \left(a^2+b^2\right)+ \gamma _- \left(a^2-b^2\right)+U_2 \right) +\Gamma a b  d&=0\\
d\left(\beta _+ \left(c^2+d^2\right)-\beta _- \left(c^2-d^2\right)+\gamma _+ \left(a^2+b^2\right)- \gamma _- \left(a^2-b^2\right)+U_2 \right)+\Gamma a b c &=0 \\
v^\prime\left(4 \sqrt{3} \lambda _1 v^{\prime2}+3 \rho _1 v^\prime +U_3\right)&=0,
\end{align}
\end{subequations}
where the equations have been rescaled to eliminate constant overall factors and with the shorthand notations
\begin{align*}
U_i&=\frac12 \left(\mu _i^2+\sqrt{3} \zeta _{i3}\, v^{\prime2}\right) \quad \mathrm{for}\quad i=1,2\;, & U_3 &=2 \mu_3^2+\zeta_{13}(a^2+b^2)+\zeta_{23} (c^2+d^2)
\end{align*}
and 
\begin{align*}
\left\{\begin{array}{c}
\xi_+=\frac{\xi_1}{2},\quad \xi_-=\frac{\xi_2+\xi_3}{2\sqrt{3}}\quad\mathrm{for}\quad \xi=\alpha,\beta\\
\gamma_+=\frac{\sqrt{3}\gamma_1+\gamma_4}{4\sqrt{3}},\quad\gamma_-=\frac{\gamma_2+\gamma_3}{4\sqrt{3}}\quad\mathrm{and}\quad\Gamma=\frac{ \gamma_4}{\sqrt{3}}
\end{array}\right\}
&\;\mathrm{for}\, \ev{S}
\\
\left\{\begin{array}{c}
\xi_+=\frac{\sqrt{3}\xi_1+\xi_2+\xi_3}{2\sqrt{3}},\quad \xi_-= \frac{2\xi_3-\xi_2}{2\sqrt{3}}\quad\mathrm{for}\quad
\xi=\alpha, \beta,\\
\gamma _+=\frac{\sqrt{3}\gamma_1+\gamma_3}{4\sqrt{3}},
   \quad\gamma _-=\frac{\gamma_3}{2\sqrt{3}}
   \quad\mathrm{and}\quad\Gamma=\frac{\gamma_2+\gamma_4}{2\sqrt{3}}
   \end{array}\right\}
&\;\mathrm{for}\, \ev{SY}\\
\left\{\begin{array}{c}
\xi_+=\frac{\xi_1+\sqrt{3}\xi_2}{2},\quad
\xi_-= \frac{\xi_3-2\xi_2}{2\sqrt{3}}
\quad\mathrm{for}\quad\xi=\alpha, \beta,\\
\gamma _+=\frac{\sqrt{3}\gamma_1+\gamma_2}{4\sqrt{3}},\quad
\gamma_-=\frac{\gamma_3+\gamma_4}{4\sqrt{3}}
\quad\mathrm{and}\quad\Gamma=\frac{\gamma_2}{\sqrt{3}}
\end{array}\right\}
&\;\mathrm{for}\, \ev{SYX}\;.
\end{align*}
The first four equations result from the derivatives taken with respect to the
components of $\phi_1$ and $\phi_2$. The eleven minimization conditions, corresponding to the 11 real scalar degrees of freedom, thus reduce to just five equations for five unknowns and there generally is a solution. Thus, there is no vacuum alignment problem in this model. We have checked numerically that this is the global minimum for a region of parameter space.

Note that the equations for $v^\prime$ and $a,b,c,d$ essentially decouple and the contribution to the other one, can be reabsorbed in the mass term. They are invariant under symmetries $(a,c)\leftrightarrow (b,d)$,  $(a,b)\rightarrow -(a,b)$, $(c,d)\rightarrow-(c,d)$ as well as $(a,b,\alpha_i,U_1)\leftrightarrow (c,d,\beta_i,U_2)$, which are inherited from the symmetries of the potential.  

There are also minima breaking all symmetries.
Generally, for each minimum, there are $95$ additional minima with the same value $V\vert_\mathrm{min}$, which are connected by a group transformation. This multitude of minima makes an analytic treatment unfeasible. Therefore, we have performed a numerical study. We have varied all parameters in the range $[ -4,4]$ and only found minima corresponding to the ones given above. In particular, we have not found any minima that leave a larger symmetry group intact (except for the minimum with vanishing VEVs, which does not break the group). Each of these minima can be realised as global minimum of the potential, which we checked in the random number scan we performed. However, it was impracticable to determine the parameter regions of each global minimum. 


\section{Higher Order Corrections}
\label{sec:higherorderop}
The results presented above are corrected by higher order operators. Here we discuss the next-to-leading order corrections.
Let us briefly comment on the magnitude of the scale $\Lambda$ under the assumption that all operators are suppressed by the same scale.\footnote{Of course, this assumption does not have to be true for e.g. a UV completion where the charged lepton mass operators are generated by vector-like fermions and the neutrino mass operators are generated by a see-saw.} If we require a perturbative value for the $\tau$ Yukawa coupling, $y_\tau<4 \pi$, this translates into~\cite{Altarelli:2005yx}
$$\frac{v'}{\Lambda}>0.002.$$ 
Furthermore taking $m_\nu\sim0.05\eV$ and assuming couplings of order one, $x_{a,d}\sim \Order{1}$, we find
$$\Lambda\approx 6\cdot 10^{14}\left( \frac{u}{\Lambda} \right)^2\GeV\quad\mathrm{with}\quad u=a,b,c,d$$
and the natural cutoff values are therefore $2 \cdot 10^9
\GeV<\Lambda<6\cdot 10^{14}\GeV$ assuming all VEVs to be of a
similar size $v^\prime\sim a\sim b\sim\dots$, but it can easily be in the TeV region for moderately small couplings in the UV completion. 

\subsection{Corrections to the charged lepton mass matrix}
The next-to-leading order correction to the charged lepton mass matrix takes the form:
\begin{align}
\mathcal{L}_e^{( 6)} = y_e' (\ell (\chi\chi)_{\MoreRep{3}{1}} )_{\MoreRep{1}{1}} e^c  \tilde{H}/\Lambda^2
+y_\mu' (\ell (\chi\chi)_{\MoreRep{3}{1}})_{\MoreRep{1}{3}} \mu^c  \tilde{H}/\Lambda^2
+y_\tau' (\ell (\chi\chi)_{\MoreRep{3}{1}})_{\MoreRep{1}{2}} \tau^c  \tilde{H}/\Lambda^2 +\hc\;.
\end{align}
As these operators can be obtained by replacing $\chi$ by $(\chi\chi)_{\MoreRep{3}{1}}$ in Eq. (\ref{eq:chargedlepton}) and 
$$\langle (\chi\chi)_{\MoreRep{3}{1}} \rangle=v^\prime \langle \chi \rangle,$$
they do not introduce a new structure in the charged lepton mass matrix~\cite{Altarelli:2006kg}, but merely renormalise the leading contribution. Note that there are no other contributions at this level, since $\phi_i\phi_i$ does not contain \MoreRep{3}{1} by construction. Operators with new structures are suppressed by $1/\Lambda^3$.
\subsection{Corrections to the neutrino mass matrix}
The next-to-leading order operators contributing to the neutrino mass matrix are given by
\begin{multline}
\Lambda^4\mathcal{L}_\nu^{(8)}= x_c (\ell H\ell H)_{\MoreRep{1}{2}} (\phi_1 \phi_2 \chi)_{\MoreRep{1}{3}} +x_b (\ell H\ell H)_{\MoreRep{1}{3}} (\phi_1 \phi_2 \chi)_{\MoreRep{1}{2}} +x_h (\ell H\ell H)_{\MoreRep{1}{1}} (\phi_1 \phi_2 \chi)_{\MoreRep{1}{1}}+\\
+(\ell H\ell H)_{\MoreRep{3}{1}}\cdot\left[x_e  \chi   (\phi_1 \phi_2 )_{\MoreRep{1}{1}}
+x_f (\chi\cdot (\phi_1 \phi_2 )_{\MoreRep{3}{1}})_{S} 
+x_g (\chi\cdot (\phi_1 \phi_2 )_{\MoreRep{3}{1}})_{A} \right]+\hc \;,
\label{eq:neutrinomassop9}
\end{multline}
where $(\dots)_S$ denotes the symmetric contraction and $(\dots)_A$ the antisymmetric one.
These operators perturb the mixing matrix and their effect will be discussed in section \ref{sec:masses-mixings}.
\subsection{Corrections to the Scalar Potential}
 Corrections to the potential arise at dimension five:
\begin{multline}
V^{(5)}=\sum_{L,M=1}^{2}\sum_{i,j=2}^{4} \frac{\delta^{(LM)}_{ij}} {\Lambda} \chi\cdot\left\{ (\phi_L \phi_L) _{\MoreRep{3}{i}} \cdot (\phi_M \phi_M) _{\MoreRep{3}{j}}\right \}_{\MoreRep{3}{1}}+\\
+\frac{\chi^3}{\Lambda}\left( \delta^{(3)}_1\chi^2+\delta^{(3)}_2 (\phi_1 \phi_1) _{\MoreRep{1}{1}}+\delta^{(3)}_3 (\phi_2 \phi_2) _{\MoreRep{1}{1}} \right)
\label{eq:V5}
\end{multline}
where all parameters are real and $\delta^{(LM)}_{ij}=0$ for $i \geq j$.
Upon minimisation, these interactions lead to a shift in the vacuum expectation values of the form:
\begin{subequations}
\begin{align}
\langle\chi\rangle &=(v'+\delta v_1',v'+\delta v_2',v'+\delta v'_2)^T,\\
\langle\phi_1\rangle &=\frac{1}{\sqrt{2}}(a+\delta a_1 ,a+\delta a_2,b+\delta a_3,-b+\delta a_4)^T,\\
\langle\phi_2\rangle &=\frac{1}{\sqrt{2}}(c+\delta b_1,c+\delta
b_2,d+\delta b_3,-d+\delta b_4)^T
\end{align}
\end{subequations}
Generically, the magnitude of these shifts will be suppressed by one power of $\Lambda$,
\begin{align}
\frac{\delta u}{u}\sim\frac{u}{\Lambda},
\end{align}
where u denotes a generic vacuum expectation value. The VEVs of $\chi_2$ and $\chi_3$ stay equal at next-to-leading order,  i.e. $\ev{\chi_2}-\ev{\chi_3}=\mathcal{O}(1/\Lambda^2)$ and $\ev{\chi_3}\approx\ev{\chi_2}=\delta v_2^\prime$. To calculate the correction to neutrino masses, the following shorthand notations for the shifts in the vacuum expectation values are useful:
\begin{align}
\delta \langle  (\phi_1 \phi_2) _{\MoreRep{3}{1}} \rangle=\frac{1}{4} \left(\begin{array}{c}
 a \left(\delta b_4 -\delta b_3 \right)+c \left(\delta a_3 -\delta a_4
   \right)-d \left(\delta a_1 +\delta a_2 \right)+b \left(\delta b_1 +\delta b_2
   \right)\\
 a \left(\delta b_3 +\delta b_4 \right)-c \left(\delta a_3 +\delta a_4
   \right)+d \left(\delta a_1 -\delta a_2 \right)+b \left(\delta b_2 -\delta b_1
   \right)\\
a \left(\delta b_1 -\delta b_2 \right)+c \left(\delta a_2 -\delta a_1
   \right)-d \left(\delta a_3 +\delta a_4 \right)+b \left(\delta b_3 +\delta b_4
   \right)
\end{array}\right)\equiv \left( \begin{array}{c}
 \delta \Phi_1\\ \delta \Phi_2 \\ \delta \Phi_3
 \end{array}\right)
\end{align}
and 
\begin{align}
\delta \langle  (\phi_1 \phi_2) _{\MoreRep{1}{1}} \rangle=\frac{1}{4} \left(a \left(\delta b_1 +\delta b_2 \right)+c \left(\delta a_1 +\delta a_2
   \right)+d \left(\delta a_3 -\delta a_4 \right)+b \left(\delta b_3 -\delta b_4
   \right)\right)\equiv \delta\Phi_0\;.
\end{align}

\begin{figure}[tb]
\centering
\includegraphics[width=.7\textwidth]{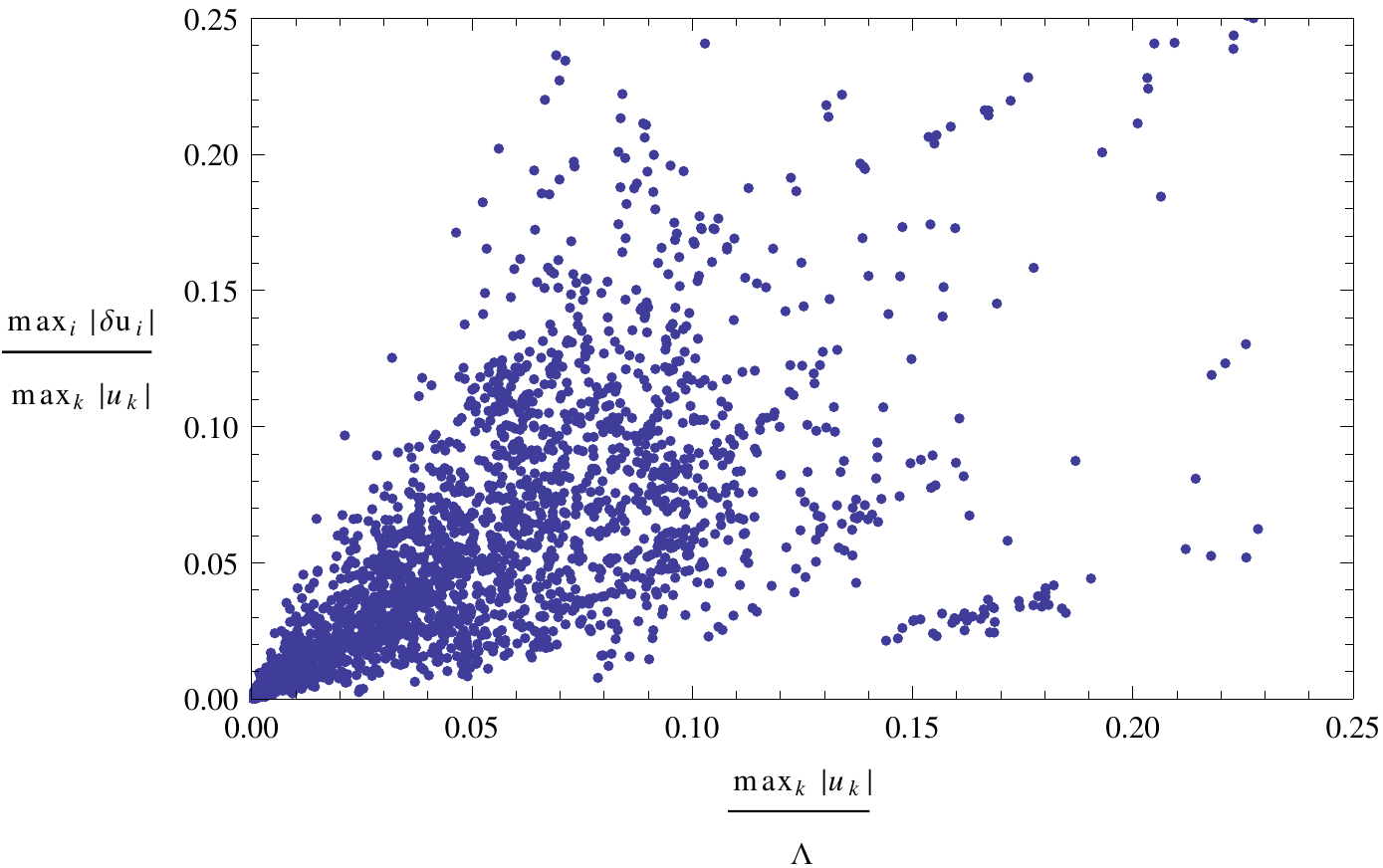} 
\caption{Vacuum shifts $\frac{\max_i |\delta u_i|}{\max_k |u_k|}$ induced by higher dimensional operators as a function of $\frac{\max_k |u_k|}{\Lambda} $ for randomly chosen potential parameters of order unity. All points correspond to phenomenologically viable data points. }\label{fig:vevs}
\end{figure}
To get a feeling for the size of the deviations from the leading order vacuum alignment, we have performed a numerical minimisation of the potential for a number of random values for the potential parameters. We found it instructive to plot $\frac{\max_i \delta u_i}{\max_i u_i}$ against $\frac{\max_i u_i}{\Lambda} $, where $u_i$ denotes any of the leading-order VEVs and $\delta u_i$ any of the deviations. \Figref{fig:vevs} shows the VEV deviation scales plotted against the ratio $u/\Lambda$. The corrections are small for small $u/\Lambda$.

\subsection{Corrections of Masses and Mixings}
\label{sec:masses-mixings}
To next-to-leading order, the charged lepton matrix $M_E$ is modified from \Eqref{eq:numass} by 
\begin{align}
\delta M_E &=\frac{v }{\Lambda \sqrt{2}} \left( \begin{array}{ccc} \delta v^\prime_1 &0&0\\0& \delta v^\prime_2  & 0  \\0&0& \delta v^\prime_2 \end{array}\right)  U_0 \, \left( \begin{array}{ccc} y_e&0&0\\0& y_\mu & 0 \\0& 0&y_\tau \end{array}\right)+\frac{v {v^\prime}^2}{\Lambda^2 \sqrt{2}}  U_0 \, \left( \begin{array}{ccc} y_e^\prime &0&0\\0& y_\mu^\prime & 0 \\0& 0&y_\tau^\prime \end{array}\right)\;.
\end{align}
In the neutrino sector there are also new structures. The corrections to the neutrino mass matrix can be parametrised as
\begin{align}
\delta m_\nu=\left(\begin{array}{ccc}
\delta \tilde{a}+\tilde{b}+\tilde{c}&\tilde{f}&\tilde{e}\\
\tilde{f}&\delta \tilde{a}+\omega \tilde{b}+\omega^2\tilde{c}&\delta \tilde{d}\\
\tilde{e}&\delta \tilde{d}&\delta\tilde{a}+\omega^2\tilde{b}+\omega \tilde{c}
\end{array}\right)\frac{v^2}{2}
\end{align}
with 
\begin{subequations}
\begin{align}
\delta \tilde{a}&=\frac{v' x_{h} (b c-a d)}{6 \Lambda ^4}+\frac{x_{a } \delta \Phi_0}{
   \sqrt{3} \Lambda ^3}\;,&
 \delta \tilde{d}&=\frac{-x_d \delta \Phi_1}{2 \sqrt{3} \Lambda ^3}+\frac{v' x_{e} (ac+bd)}{4 \sqrt{3} \Lambda ^4},
 \end{align}
 \begin{align}
   \tilde{b}&= \frac{v' x_b (b c-a d)}{6 \Lambda ^4}\;,&
  \tilde{e}&=\frac{-x_d \delta \Phi_2}{2 \sqrt{3} \Lambda ^3}+\frac{v' x_{e} (ac+bd)}{4 \sqrt{3} \Lambda ^4}+\frac{(x_f+x_g) v'  (b c-a d)}{8 \sqrt{3} \Lambda ^4}\;,\\
 \tilde{c}&=\frac{v' x_c (b c-a d)}{6 \Lambda ^4}\;,&
  \tilde{f}&=\frac{-x_d \delta \Phi_3}{2 \sqrt{3} \Lambda ^3}+\frac{v' x_{e} (ac+bd)}{4 \sqrt{3} \Lambda ^4}+\frac{(x_f-x_g) v'  (b c-a d)}{8 \sqrt{3} \Lambda ^4}\;.
\end{align}
\end{subequations}

As the leptons only transform under the $A_4$ subgroup of the model, the neutrino phenomenology runs exactly parallel to the $A_4$ case. The effects of the operators $\tilde{a},...,\tilde{f}$ have been studied in \cite{Honda:2008rs} where it has been shown that a sizeable deviation from $\sin^2 \theta_{13}=0$ is possible without introducing large corrections to the other mixing angles.
Recently it has been shown that $\sin^2 \theta_{13}\gtrsim 0.1$ is possible for $\tilde{c}/\tilde{a}\gtrsim 0.25$ in the case of normal mass ordering~\cite{Shimizu:2011xg}.

\begin{figure}[tb]
  \centering
  \subfigure[$\sin^2 \theta_{12}$ vs.~$\sin^2 \theta_{13}$ ]{
    \label{fig:12-13}
    \includegraphics[width=.45\textwidth]{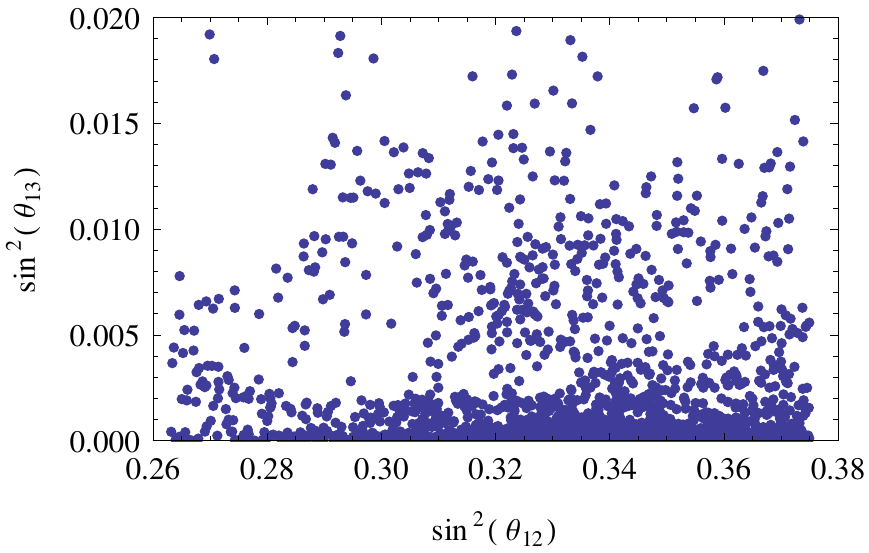} 
  }
  \subfigure[$\sin^2 \theta_{12}$ vs.~$\sin^2 \theta_{23}$]{
    \label{fig:12-23}
    \includegraphics[width=.45\textwidth] {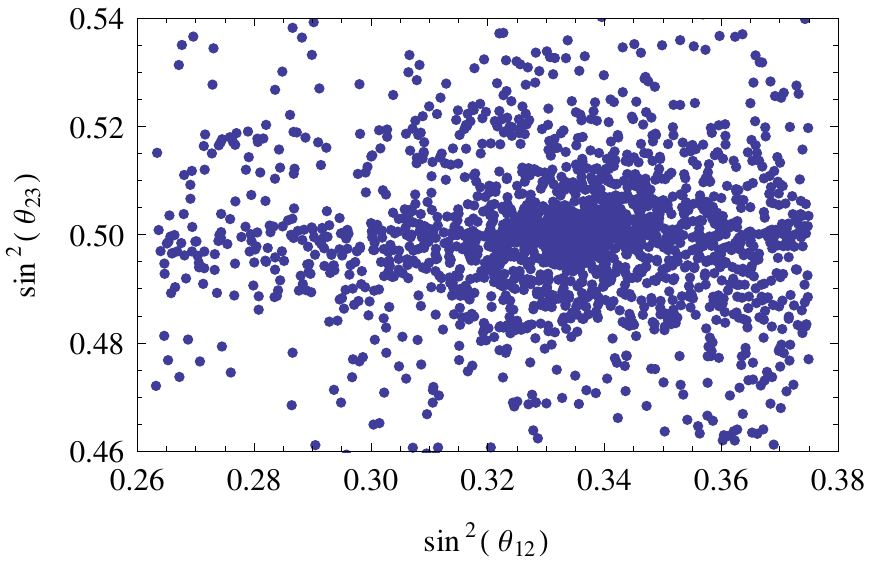} 
  }
    \subfigure[$\sin^2 \theta_{13}$ vs.~$\sin^2 \theta_{23}$]{
    \label{fig:13-23}
    \includegraphics[width=.45\textwidth] {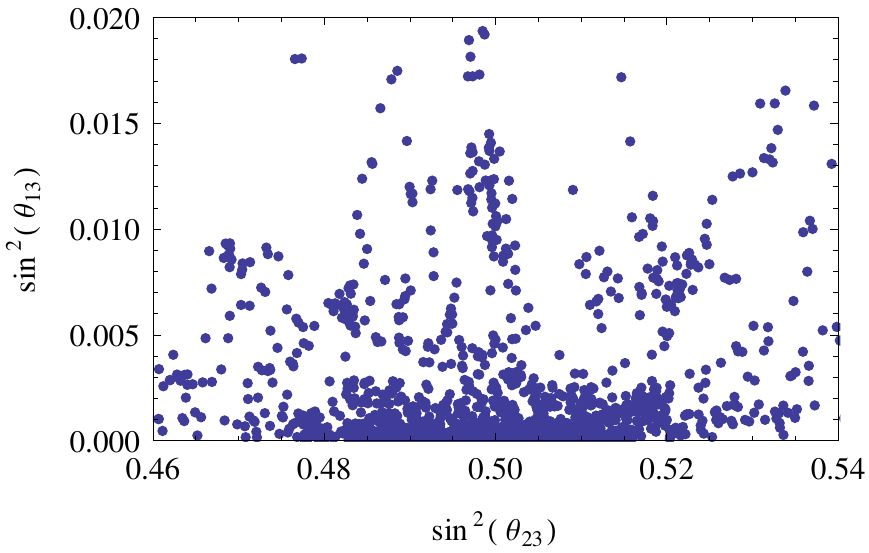} 
  }
      \subfigure[$\sin^2 \theta_{13}$ vs.~$\frac{\max_i u_i}{\Lambda} $]{
    \label{fig:13-23}
    \includegraphics[width=.45\textwidth] {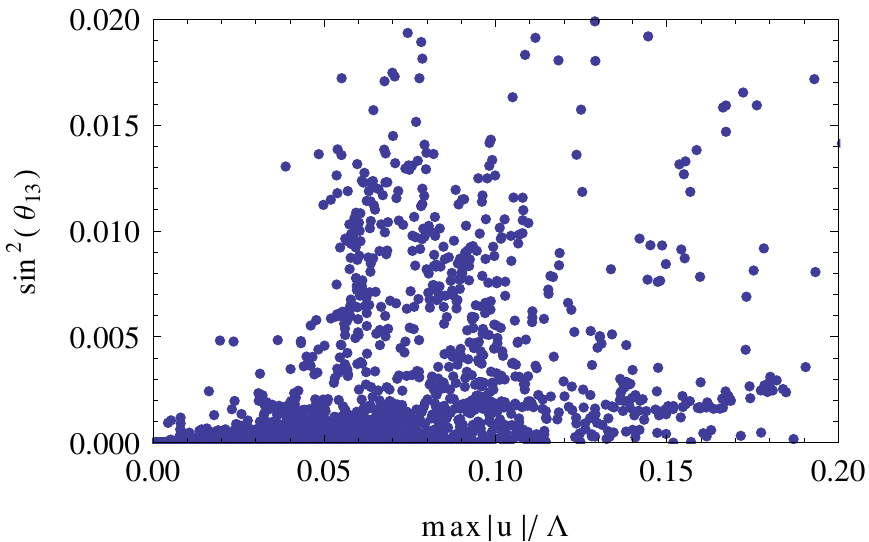} 
  }
  \caption{\textbf{Scatterplot of Mixing Angles.} To illustrate the typical size of corrections to the mixing angles, we have performed a scatter plot. We took all dimensionless scalar potential couplings to be of order one and varied the ratio of the mass parameters in the potential such that the ratio of the VEVs and cutoff- scale is smaller than one. All dimensionless parameters that modify the neutrino and charged lepton mass matrices are taken of the same order as the leading order parameters. All points lie within the 3 $\sigma$ range mass of the mass and mixing parameters. The mixing angle $\sin^2\theta_{12}$ is varied more than the the other two mixing angles.  We have used the \texttt{MixingParameterTools}~\cite{Antusch:2005gp} package to extract the mixing angles.
  }
  \label{fig:scatter}
\end{figure}

We performed a scatter plot in order to get an idea of the size of the corrections from higher dimensional operators.
For a collection of tree-level parameters of order unity, we have varied the higher dimensional parameters (\ref{eq:V5}) of the potential in the range $[0.5,1.5]$ and the dimensionless parameters in the corrections to the lepton masses in \Eqref{eq:neutrinomassop9} have been taken to be of the same order as the leading order contributions. The suppression scale $\Lambda$ has been varied in a wide range.

In \Figref{fig:scatter}, the resulting scatter plots are shown, where all data points lie within the $3\,\sigma$ limits of the global fits cited in the introduction. As can be seen from \Figref{fig:13-23}, for $u/\Lambda\gtrsim 0.05$ there are points that deviate from tribimaximal mixing in the right way to be compatible with the recent measurements from T2K. 

\begin{figure}[tb]
  \centering
    \includegraphics[width=.7\textwidth]{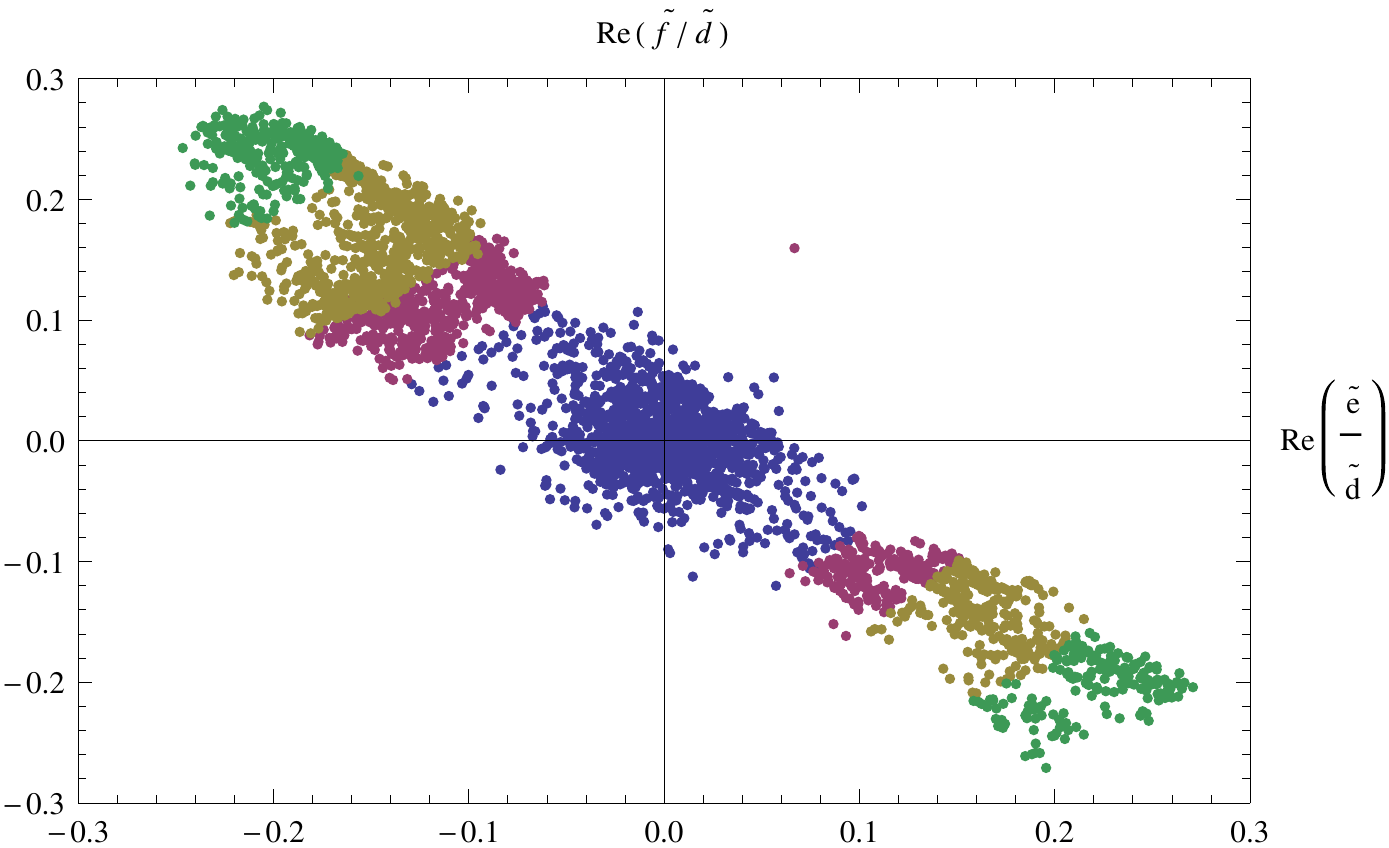}
  \caption{The matrix entries $\tilde{e}$ and $\tilde{f}$ are the dominant corrections to the neutrino mixing matrix in the case where the corrections from the VEVs of $\phi_1$ and $\phi_2$ dominate. Here we show the correlation between the two quantities. The blue, violet, yellow and green points correspond to values of $\sin\theta_{13}^2$ in the ranges $[0,0.005]$, $[0.005,0.01]$,  $[0.01,0.02]$, and $[0.02,0.03]$, respectively. The imaginary parts are much smaller $\left|\mathrm{im}\left({\tilde{e}}/{\tilde{d}}\right)\right|, \left|\mathrm{im}\left({\tilde{f}}/{\tilde{d}}\right)\right|\lesssim 0.003$.}
  \label{fig:c/a}
\end{figure}

Allowing for couplings considerably smaller than order one in $V_\phi$\footnote{For details, please consult the Mathematica notebook published as a supplement together with the Mathematica package \Discrete\  described in \Secref{sec:Mathematica}}, the VEV corrections $\delta \Phi_i$ become dominant and $\tilde{e}$ and $\tilde{f}$ are the main corrections to the neutrino mass matrix. This is shown in Figure \ref{fig:c/a} and is in agreement with the result \cite{Honda:2008rs}. Note that the values are roughly along a diagonal line, i.e. $\tilde{e}$ and $\tilde{f}$ are similar in size, but have a different relative sign.

\subsection{Cosmological Implications of Accidental Symmetries}
\label{sec:cosmo-impl}
Let us briefly comment on possible cosmological implications of the unbroken remnant $Z_2$ and $Z_3$ symmetries in the neutrino and charged lepton sectors, respectively. Due to the accidental $A_4$ symmetry in the scalar potential these symmetries are accidental symmetries of the theory only broken by higher dimensional operators. 

Let us discuss the situation where $J$ is the lightest scalar odd under the unbroken $Z_2$ symmetry generated by $S$, e.g. $J=\frac{1}{\sqrt{2}} \left((\phi_1)_3+(\phi_1)_4 \right)$. It can then decay into neutrinos through the effective interaction 
\begin{align}
\mathcal{L}=-\frac 12 g_{J\nu_i \nu_j} J \nu_i \nu_j +\hc
\end{align}
with a lifetime roughly given by
\begin{align}
\tau(J\rightarrow\nu \nu)\sim \frac{16 \pi}{m_J}\frac{u^2}{m_{\nu} ^2}\sim 4\cdot 10^{8}\, \mathrm{s}\,\left(\frac{u}{m_J}\right)\left(\frac{u}{10^{10}\GeV}\right) 
\end{align}
for $m_\nu=0.05 \eV$ and $u$ being a generic flavon VEV . Depending on the model parameters, this decay time can be problematic. If the lifetime is larger than the age of the Universe, $J$ becomes a dark matter candidate. A large lifetime naturally occurs, if $J$ is a pseudo-Goldstone boson~\cite{Lattanzi:2007ux}, which leads to $m_J/u\ll 1$. Pseudo-Goldstone bosons often appear in these constructions. For example, the tree-level scalar potential of the next-larger group in \Tabref{tab:candidates}, $T^\prime \rtimes A_4$, has the large continuous accidental symmetry $\mathrm{Sp}(4)$.

However, in general, there is also the decay channel via higher dimensional operators in the scalar potential, which couple $J$ to the $\ev{S}$-breaking VEV of $\chi$, e.g.~by operators of the type $\phi_1^4\cdot \chi H^{\dagger}H$.  It will generically be the dominant decay process in the model outlined above and result in much shorter lifetimes of 
$$
\tau \sim 16 \pi \frac{m_J \Lambda^6}{u^8}\sim 3.3  \cdot 10^{-21}\, \left(\frac{m_J}{u}\right)\left(\frac{u/\Lambda}{0.01}\right)^{-7}\left(\frac{10^{12}\GeV}{\Lambda}\right)\,\mathrm{s}
$$
ensuring that any potential abundance of $J$ will decay before big bang nucleosynthesis.
In the model by Babu and Gabriel~\cite{Babu:2010bx}, these higher dimensional operators are absent and therefore this decay through neutrinos is the only decay channel, which poses a potential problem for these models.

\section{See-saw UV completion}
\label{sec:UV-completion}
The neutrino sector of the effective theory outlined above, may be UV
completed by introducing the left-handed Weyl spinors $N$, $S_2$ and
$S_3$ that transform under $(Q_8\rtimes A_4)\times Z_4$ as $N\sim
(\MoreRep{3}{1},-\I)$, $S_2\sim (\MoreRep{4}{2},\I)$ and $S_3\sim (\MoreRep{4}{3},-\I)$. $S_2$ and $S_3$ can be combined in a Dirac spinor.

\begin{figure}[tb]
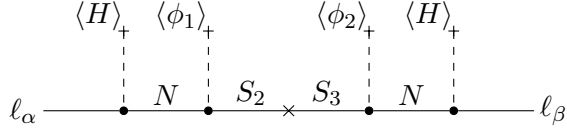

  \centering
     \FMDG{DS2}
  \caption{Neutrino masses in the UV completion.}
  \label{fig:UV}
\end{figure}

This leads to the following new interactions in the Lagrangian
\begin{multline}
\mathcal{L}=x_{\ell N} \ell H N +x_{N2} N S_2 \phi_1  +x_{N3} N S_3 \phi_2+ m
S_2S_3 +x_{23} S_2 S_3 \chi+\hc\;,
\end{multline}
where the contraction of each operator is uniquely determined by the group theory of $Q_8\rtimes A_4$.
The neutral fermion mass matrix is then schematically given by
\begin{equation}
\frac12\left(\begin{array}{ccccc}
0 & x_{\ell N} \ev{H}  & 0 & 0\\
\dots & 0  & x_{N2} \ev{\phi_1} & x_{N3} \ev{\phi_2}\\
\dots & \dots  & 0 & m + x_{23} \ev{\chi}\\
\dots & \dots  & \dots &0 \\
\end{array}\right)
\end{equation}
in the basis $\left(\nu,\,N,\,S_2,\,S_3\right)$. In the following, we assume that the direct mass term is larger than the mass terms generated by VEVs. Therefore, we are in the seesaw regime, which has been firstly studied for gauge singlets in~\cite{Minkowski:1977sc,*Yanagida:1980,*Glashow:1979vf,*Gell-Mann:1980vs,*Mohapatra:1980ia} and in more generality in \cite{PhysRevD.22.2227,*PhysRevD.25.774}. 
Hence, the masses of the singlets $N$ are generated
\begin{equation}
m_{N}=\frac{x_{N2}x_{N3}}{m}\left(\begin{array}{ccc}
A&0&0\\0&A&B\\0&B&A
\end{array} \right)
\quad\mathrm{with }\quad
A=-2(a c +b d )\quad\mathrm{and}\quad
B=\I \sqrt{3} (b c -a d)\;.
\end{equation}
This particular form has been denoted linear see-saw~\cite{Malinsky:2005bi}. The light neutrino masses are generated via a standard see-saw~\cite{Minkowski:1977sc,*Yanagida:1980,*Glashow:1979vf,*Gell-Mann:1980vs,*Mohapatra:1980ia}.
Hence, the operator $x_{23} S_2S_3\chi$ does only enter at next-to leading order. Alternatively, it is possible to forbid it together with all next-to leading order corrections, which have been discussed in the previous section, by introducing an additional $Z_2$ symmetry $\chi\rightarrow -\chi$ and $f^c\rightarrow -f^c$ with $f$ being $e$, $\mu$ or $\tau$. The neutrino mass matrix is then given by
\begin{align}
m_\nu=x_{\ell N}^2 v^2 m_N^{-1} 
\end{align}
This can also be seen from Figure \ref{fig:UV}. This matrix is
diagonalized by $U_\nu$:  $U_{\nu}^T  m_\nu
U_{\nu}=\diag(\frac{1}{B+A},\frac{1}{A},\frac{1}{B-A})$. However,
there are two degenerate eigenvalues as the relative phase of $A$ and
$B$ is given by $\pi/2$. This can be solved by adding another copy of
$S_2$ or $S_3$, for example, lifting the degeneracy.

The charged lepton mass operators can be generated in the same way as in \cite{Babu:2010bx} by introducing additional states that have masses allowed by EW symmetry and mix with the SM states after EW symmetry breaking.
\section{Supersymmetrization}
\label{sec:SUSY-Model}
Supersymmetrization of the model is rather straightforward. One only has to ensure that there are no flat directions in the potential:
\begin{align}
V=V_{\mathrm{SUSY}}+V_{\mathrm{soft}}
\end{align}
with 
\begin{align}
V_{\mathrm{SUSY}}=\sum_{i}\abs{ \frac{\partial W}{\partial \varphi_i} }^2
\end{align}
where W denotes the superpotential and $\varphi_i$ is any of the fields in Table \ref{tab:SUSYpartcontent}. $V_{\mathrm{soft}}$  contains all supersymmetry-breaking soft terms invariant under the flavour symmetry. 
\begin{table}
\begin{center}
\begin{tabular}{|l|ccc||c|c|}\hline
particle & ${SU}({3})_c$ & ${SU}({2})_L$ & ${U}({1})_Y$&
$Q_8\rtimes A_4$&$Z_{4}$\\\hline\hline
$\ell$ & 1 & 2 & -1/2  &\MoreRep{3}{1}&$\I$\\
$e^c+\mu^c+\tau^c$ & 1 & 1 & 1   & $\MoreRep{1}{1}+\MoreRep{1}{2}+\MoreRep{1}{3}$&$-\I$ \\\hline
$H_u$ & 1 & 2 & 1/2  &\MoreRep{1}{1} &1 \\\hline
$H_d$ & 1 & 2 & -1/2  &\MoreRep{1}{1} &1 \\\hline
$\chi$ & 1 & 1 & 0  &\MoreRep{3}{1} &1 \\\hline
$\tilde{\chi}$ & 1 & 1 & 0  &\MoreRep{3}{2} &1 \\\hline
$\phi_1$ & 1 & 1 & 0  &\MoreRep{4}{1} &1\\
$\phi_2$ & 1 & 1 & 0  &\MoreRep{4}{1}&-1 \\\hline
$S$ & 1 & 1 & 0  &\MoreRep{1}{1}&1  \\\hline
\end{tabular}
\caption{Chiral Superfield Particle Content.\label{tab:SUSYpartcontent}}
\end{center}
\end{table}

As there is no cubic invariant containing the $\phi_{1,2}$ fields only  and the quadratic term $\phi_{1,2}^2=\sum_{i}{\phi_{1,2}^{i}}^2$ is $\SO{4}^2$ invariant under the individual rotations of $\phi_1$ and $\phi_2$, we have to add the singlet $S$ and the triplet $\tilde{\chi}\sim \MoreRep{3}{2}$, to get a superpotential without flat directions in the cubic terms and without a continuous accidental symmetry. We thus have the schematic superpotential 
\begin{align}
W&=S	(\phi_1^2+\phi_2^2+\chi^2+\tilde{\chi}^2)_{\MoreRep{1}{1}}+ S^3+ S^2+S+\phi_1^2+\phi_2^2 +\chi^2+\chi^3+\tilde{\chi}(\phi_1^2+\phi_2^2)_{\MoreRep{3}{2}}+\tilde{\chi}^2+\tilde{\chi}^3.
\end{align}
We have studied the potential resulting from this superpotential and the most general soft-breaking terms and we have found a portion of parameter space with the right vacuum alignment, with non-vanishing VEVs for both the singlet and triplet contractions of the product $\phi_1 \phi_2$. The neutrino mass operators are again given by
\begin{align}
W &\supset x_{a} (\ell H\ell H)_{\MoreRep{1}{1}} (\phi_1 \phi_2)_{\MoreRep{1}{1}} /\Lambda^3 + x_d (\ell H_u\ell H_u)_{\MoreRep{3}{1}} (\phi_1 \phi_2)_{\MoreRep{3}{1}} /\Lambda^3  \label{eq:neutrinomassop}.
\end{align}
As in the non-SUSY model before, the the on-and off-diagonal terms of the neutrino mass matrix, which have to be quite close to each other in magnitude, are generated by VEVs of the same fields. 
The additional scalar field $\tilde{\chi}$ couples to leptons only on next-to next-to leading order and it is thus not problematic. 

Details can be found in the Mathematica notebook accompanying this
paper, which can be downloaded from the \href{http://projects.hepforge.org/discrete}{webpage} of the Mathematica
package \Discrete. We do not give the details here, as it is somewhat out of the main
focus of the paper, but we have checked that there exist parameter values for which the global minimum of the potential has the correct vacuum alignment for the most general softly broken supersymmetric potential. The symmetry breaking is also complete, i.e. there are no flat directions left as is the case in the model by Altarelli and Feruglio~\cite{Altarelli:2006kg}, where one introduces driving fields with $U_R(1)$ charges of 2. It has been shown that the inclusion of soft-breaking terms in this model is problematic, as it generically leads to flavour violating VEVs of auxiliary fields~\cite{Feruglio:2009iu}, unless there is a solution to the SUSY flavour problem in terms of gauge mediation~\cite{Giudice:1998bp} or via a separate mechanism (see e.g. \cite{Antusch:2008jf}).

It would be interesting to come back to the SUSY VEV alignment problem and search for groups that do not need additional scalar fields to break accidental symmetries.


\section{\Discrete\ --- Mathematica Package}
\label{sec:Mathematica}

\Discrete\ is a Mathematica package with several useful model building tools to work with discrete symmetries. The main features are
\begin{itemize}
\item the calculation of arbitrary Kronecker products,
\item an interface to the group catalogues within \GAP~\cite{GAP4:2011}, e.g.~the \SmallGroups~\cite{SmallGroups:2011} library with all discrete groups up to order $2000$ (with the exception of groups of order $1024$) and many more. 
\item calculation of Clebsch-Gordan coefficients. They are calculated on demand and are stored internally, in order to improve the performance.
\item the possibility to reduce covariants to a smaller set of independent covariants.
\item the documentation is integrated in the documentation centre of
Mathematica. 
\end{itemize}
\Discrete\ can be downloaded from
\url{http://projects.hepforge.org/discrete/}. It has been tested with
Mathematica 8 running on Linux as well as MacOS, but we expect it to
run on older versions of Mathematica as well.

It requires a working installation of \GAP~\cite{GAP4:2011} as well as the \GAP\ package \Repsn~\cite{REPSN:2011}. \GAP\ including all its packages can be downloaded from \url{http://www.gap-system.org/}. On Debian-based Linux-distributions, it can be directly installed via the package management. 

In the following, we are presenting a short example of the abilities of \Discrete\ and refer the interested reader to the documentation and the example notebook within the package. For simplicity, we are choosing $A_4$ and only calculate the renormalizable part of the flavon potential of $\phi\sim\Rep{3}$. For brevity, we have shortened the output. The omissions are denoted by dots. 

\vspace{2ex}

\begin{minipage}{\textwidth}
\noindent {\bf Needs["Discrete$\grave{}$ModelBuildingTools$\grave{}$"];}

\noindent {\bf A4=MBloadGAPGroup["AlternatingGroup(4)"];}

\noindent{\dots}

\noindent {Dimensions of irreps:\\
$\begin{array}{llll}
 1 & 2 & 3 & 4 \\
 1 & 1 & 1 & 3
\end{array}
$}

\noindent {\dots}

\noindent {\bf phi=MBgetRepVector[A4,4,p]}

\noindent {\bf\$Assumptions=Variables[phi]$\in$ Reals;}

\noindent {\(\{\{\},\{\},\{\},\{\{\text{p1},\text{p2},\text{p3}\}\}\}\)}

\noindent {\bf V=MBgetFlavonPotential[A4,phi,4,h]}
\begin{multline*}
\text{h3n1} \,\text{p1}\, \text{p2}\, \text{p3}+\frac{\text{h2n1} \left(\text{p1}^2+\text{p2}^2+\text{p3}^2\right)}{\sqrt{3}}+\frac{1}{3} \text{h4n2}
\left(\text{p1}^2+\text{p2}^2+\text{p3}^2\right)^2\\
+\frac{1}{3} \text{h4n1} \left(\text{p1}^4+\text{p2}^4-\text{p2}^2 \text{p3}^2+\text{p3}^4-\text{p1}^2
\left(\text{p2}^2+\text{p3}^2\right)\right)+\frac{\text{h4n3} \left(\text{p2}^2 \text{p3}^2+\text{p1}^2 \left(\text{p2}^2+\text{p3}^2\right)\right)}{\sqrt{3}}
\end{multline*}
\end{minipage}

\vspace{2ex}

After loading the package in line 1 and loading the group $A_4$ from \GAP, we define a field phi in the third line in boldface, which transforms as triplet of $A_4$. The $4$ denotes the triplet in the list of representations and the last argument determines how the components of phi are denoted. Furthermore, we declare all components of phi to be real. 
MBgetFlavonPotential returns the flavon potential of phi up to fourth order as specified in the third argument and the couplings start with h. The first number in the name of the coupling denotes the order and the second one enumerates the couplings of a given order. 

Part of the calculation for $Q_8\rtimes A_4$ is included in \Discrete\ as example. However, we recommend to start with the tutorial included in \Discrete, which introduces and explains most functions.

Recently, a Mathematica package has been presented that allows one to calculate the group invariants formed from the three dimensional representation for most finite subgroups of \SU{3} with order smaller than 512~\cite{Merle:2011kx}.


\section{Conclusions}
\label{sec:Conclusions}
In this paper, we have revisited the long-standing problem of vacuum alignment in models with a discrete flavour symmetry. In such a model, in order to obtain the correct pattern for the mixing angles, it is generally necessary to break the flavour group in a specific way to two different subgroups. This vacuum alignment, however, cannot be realized as a minimum of the scalar potential due to non-trivial couplings between the two sectors responsible for the breaking to the different subgroups.

We have performed a systematic scan of all discrete groups with less than 1000 group elements. For each of the flavour groups $H=A_4,\,T_7,\,S_4,\,T^\prime$ and $\Delta(27)$ of the SM fermions, we have identified a number of candidate groups $G$ with $G/N\simeq H$ and a three dimensional representation $\chi$ inherited from $H$. We further required the existence of 
a faithful representation $\phi$, such that there appears an accidental symmetry $G \times H$ in the renormalizable part of the scalar potential. The flavon $\chi$, responsible for symmetry breaking in the charged lepton sector, and the SM fields essentially only transform under the group $H$, thereby preserving the mixing angle predictions of $H$. The flavon $\phi$ breaks the symmetry in the neutrino sector and its product $\phi \times \phi$ does not contain any of the representations of $H$. This is a necessary condition for the accidental symmetry, as the term $\phi^2$ can -- infamously -- not be forbidden by an internal symmetry with unitary representations acting on $\phi$. The additional symmetry thus forbids the dangerous cross-couplings, i.e.~there is only the trivial coupling via the total singlet between $\phi$ and $\chi$ at the renormalizable level.  The accidental symmetry is then broken to $G$ by couplings to fermions and other higher dimensional interactions. 

Having identified a list of possible groups, we built an explicit model using the smallest semidirect product of the candidates, $Q_8\rtimes A_4$, as flavour group. We have used two real scalar copies $\phi_1$ and $\phi_2$ of the faithful representation $\MoreRep{4}{1}$, where the triplet contraction $ (\phi_1 \phi_2)_{\MoreRep{3}{1}}$ couples to neutrinos and thus plays the role of $\phi_S$ in the model of Altarelli and Feruglio~\cite{Altarelli:2005yx}. The accidental symmetry is protected by an additional $Z_4$ separating the charged and neutral lepton sectors. We have explicitly shown that the potential has the desired vacua and that it does not lead to unwanted pseudo-Goldstone bosons, i.e.~the symmetry breaking is complete. We have further discussed the influence of next-to-leading order higher-dimensional operators on masses and mixings.

As a direction of future work, it would be interesting to study a model where the flavons $\chi$ and $\phi_{1,2}$ transform in the same way as the Higgs field under electroweak symmetry, which would move the flavour breaking scale to the electroweak scale and make it testable. We think that this model is quite well suited for this study as there are two Higgs fields in the Weinberg operator and only one in the Yukawa couplings, as fits nicely with the structure of our model.

Let us conclude by a brief comparison with other schemes of obtaining the correct vacuum alignment. Counting degrees of freedom of the effective theory, our model has the same number of degrees of freedom as the minimal model without any mechanism for vacuum alignment~\cite{Altarelli:2005yx}. While we here do not have to add any degrees of freedom, the solution of the VEV alignment problem with an U(1)$_R$ symmetry as well as a brane constructions require a plethora of additional degrees of freedom in the form of driving fields or KK modes, respectively. When compared with the model of Babu and Gabriel based on the wreath product of $S_3$ with $A_4$~\cite{Babu:2010bx}, our model not only has a substantially lower number of degrees of freedom, but it also works as an effective theory, as $(\ell H \ell H)_{\MoreRep{3}{1}}$ and $(\ell H \ell H)_{\MoreRep{1}{1}}$ are created on the same order, which is not the case in their model.

Finally, it might be worthwhile to look into other extensions of flavour groups used in the lepton sector to address for example the quark sector. One prominent existing example is the extension of $A_4$ to $T^\prime$, which enables to describe the lepton and quark flavour structure simultaneously. We expect that our approach described in \Secref{sec:GroupScan} will be a useful tool for model building in this direction.


\section*{Acknowledgements}
We would like to thank C.~Hagedorn for discussions and comments on our draft. M.S. would like to thank K.~Petraki for discussions and would like to
acknowledge MPI f\"ur Kernphysik, where a part of this work has been done, for hospitality of its staff and the generous
support. M.H.~acknowledges support by the International Max Planck
Research School for Precision Tests of Fundamental Symmetries and thanks M. Lindner for advice. This
work was supported in part by the Australian Research Council.
We would like to thank the HepForge development environment, where the
Mathematica package \Discrete\ is currently hosted.


\appendix



\section{Clebsch-Gordan Coefficients}
\label{app:GTofQ8}
In this section, we present the Clebsch-Gordan coefficients, which are relevant for the discussion.
\subsection{\mathversion{bold}$A_4$}
The only non-trivial Kronecker product of $A_4$ is given by
\begin{equation}
\Rep{3}\times\Rep{3}=\MoreRep{1}{1}+\MoreRep{1}{2}+\MoreRep{1}{3}+\Rep{3}_S+\Rep{3}_A\;,
\end{equation}
where the indices $S$ and $A$ indicate whether the representation is
in the symmetric or antisymmetric part, respectively. The
corresponding Clebsch-Gordan coefficients, which have been computed
using \cite{PhysStatSol.b.90.211}, are
\begin{align}
\label{eq:CGcoeffA4}
(a b)_{\MoreRep{1}{1}}&= \frac{1}{\sqrt{3}}\left(  a_1 b_1 + a_2 b_2 + a_3 b_3  \right)\nonumber\\
(a b)_{\MoreRep{1}{2}}&=\frac{1}{\sqrt{3}}\left(a_1 b_1+\omega^2 a_2 b_2+\omega a_3 b_3\right)&
(a b)_{\MoreRep{1}{3}}&=\frac{1}{\sqrt{3}}\left(a_1 b_1+\omega a_2 b_2+\omega^2 a_3 b_3\right)\\\nonumber
(a b)_{A,\Rep{3}}&=\frac 12 \left(\begin{array}{c} a_2 b_3-a_3 b_2\\a_3 b_1-a_1 b_3\\a_1 b_2-a_2 b_1\end{array}\right)&
(a b)_{S,\Rep{3}}&=\frac 12\left(\begin{array}{c} a_2 b_3+a_3 b_2\\a_3 b_1+a_1 b_3\\a_1 b_2+a_2 b_1\end{array}\right)
\end{align}
where $(a_1,a_2,a_3),\,(b_1,b_2,b_3)\sim \Rep{3}$.


\begin{table}
\centering
\subtable[$G/N\cong A_4$]{\begin{tabular}{|c|cc|}
\hline
 $|G|$ & \multicolumn{2}{|c|}{\GAP}\\ \hline
$240$ & $108$& \\\hline
$288$ & $924$ &\\\hline
$336$ & $131$ &\\\hline
\multirow{3}{*}{$384$} & $618$ & $5809$ \\
& $5856$ & $18216$ \\
& $20112$ &\\\hline
$432$ & $262$ &\\\hline
$480$ & $964$ & $1041$ \\ \hline
\end{tabular}}
\hspace{2ex}
\subtable[$G/N\cong S_4$]{\begin{tabular}{|c|cccc|}
\hline
 $|G|$ & \multicolumn{4}{|c|}{\GAP}\\ \hline
$240$ & $102$ & $103$&&\\\hline
\multirow{2}{*}{$288$} & $400$ & $844$ & $845$ & $846$ \\
& $847$ & $903$&&\\\hline
$336$ & $115$ & $116$&&\\\hline
\multirow{6}{*}{$384$} & $582$ & $5614$ & $5705$ & $5708$ \\
& $5713$ & $5714$ & $5728$ & $5733$\\
& $18028$ & $18029$ & $18042$ & $18043$\\
 & $18044$ & $18045$ & $18048$ & $18102$\\
  & $18117$ & $18120$ & $18130$ & $18143$\\
   & $20069$ & $20073$&&\\\hline
$432$ & $240$ & $241$&&\\\hline
\multirow{2}{*}{$480$} & $257$ & $961$ & $967$ & $968$\\
 & $969$ & $970$ & $1020$&\\\hline\end{tabular}}
\hspace{2ex}
\subtable[$G/N\cong T^\prime$]{\begin{tabular}{|c|c|}
\hline
 $|G|$ & \GAP\\ \hline
$288$ & $409$\\\hline
$384$ & $5845$\\\hline
$480$ & $266$\\\hline
\end{tabular}}

\caption{Candidate groups $G$ of order $201-500$. $|G|$ denotes the
  order of $G$. The groups up to order $200$ are listed in \Tabref{tab:candidatesGeneral}.
Details of the groups may be accessed using the computer algebra system \GAP~by using the command
  \texttt{SmallGroup(Order,GAP)}. \label{tab:candGeneralRest}}
\end{table}

\subsection{\mathversion{bold}$Q_8\rtimes A_4$}

The product of two triplets $\MoreRep{3}{i}\times\MoreRep{3}{i}$ is described by the same Clebsch-Gordan coefficients as the one in $A_4$. They are shown in \Eqref{eq:CGcoeffA4}.
The product of two four dimensional representations $(a_1,a_2,a_3,a_4)\sim \MoreRep{4}{1}$ and $(b_1,b_2,b_3,b_4)\sim \MoreRep{4}{1}$ contains the singlet
\begin{align}
(a b)_{\MoreRep{1}{1}}&=\frac{1}{2} \left(a_1 b_1+a_2 b_2+a_3 b_3+a_4 b_4\right)
\end{align}
and the triplets:
\begin{align}
(a b)_{\MoreRep{3}{1}}&=\frac12\left(
\begin{array}{c}
-a_4 b_1+a_3 b_2-a_2 b_3+a_1 b_4\\
-a_3 b_1-a_4 b_2+a_1 b_3+a_2 b_4\\
a_2 b_1-a_1 b_2-a_4 b_3+a_3 b_4
\end{array}
\right) &
(a b)_{\MoreRep{3}{2}}&=\frac12\left(
\begin{array}{c}
a_4 b_1+a_3 b_2+a_2 b_3+a_1 b_4 \\
a_3 b_1+a_4 b_2+a_1 b_3+a_2 b_4 \\
a_2 b_1+a_1 b_2+a_4 b_3+a_3 b_4
\end{array}
\right) \nonumber\\
(a b)_{\MoreRep{3}{3}}&= \frac12\left(
\begin{array}{c}
a_1 b_1-a_2 b_2-a_3 b_3+a_4 b_4 \\
-a_1 b_1+a_2 b_2-a_3 b_3+a_4 b_4 \\
-a_1 b_1-a_2 b_2+a_3 b_3+a_4 b_4
\end{array}
\right) &
(a b)_{\MoreRep{3}{4}}&= \frac12\left(
\begin{array}{c}
a_4 b_1-a_3 b_2-a_2 b_3+a_1 b_4 \\
-a_3 b_1+a_4 b_2-a_1 b_3+a_2 b_4 \\
-a_2 b_1-a_1 b_2+a_4 b_3+a_3 b_4
\end{array}
\right) \\\nonumber
(a b)_{\MoreRep{3}{5}}&= \frac12\left(
\begin{array}{c}
-a_4 b_1-a_3 b_2+a_2 b_3+a_1 b_4 \\
a_3 b_1-a_4 b_2-a_1 b_3+a_2 b_4 \\
-a_2 b_1+a_1 b_2-a_4 b_3+a_3 b_4
\end{array}
\right)
\end{align}

\bibliography{paper}

\end{document}